\documentclass[preprint,aip,php,amsmath,amssymb]{revtex4-1}

\usepackage{bm}
\usepackage{graphicx}
\usepackage{caption,subcaption}

\usepackage{times}
\usepackage{newtxtext}

\newcommand{\wt}[1]{\widetilde{#1}}

\newcommand{\mean}[1]{\left< {#1} \right>} 
\newcommand{\tmean}[1]{\langle {#1} \rangle} 

\newcommand{\abs}[1]{{\left|#1\right|}}

\newcommand{\moment}[2]{\left< {#1}^{#2} \right>} 

\newcommand{\Ref}[1]{Ref.~\onlinecite{#1}}
\newcommand{\Refs}[1]{Refs.~\onlinecite{#1}}
\newcommand{\Eqref}[1]{Eq.~\eqref{#1}}
\newcommand{\Eqsref}[1]{Eqs.~\eqref{#1}}
\newcommand{\Figref}[1]{Fig.~\ref{#1}}
\newcommand{\Figsref}[1]{Figs.~\ref{#1}}
\newcommand{\Secref}[1]{Sec.~\ref{#1}}
\newcommand{\Appref}[1]{Appendix~\ref{#1}}

\newcommand{\Sn}{\Phi} 
\newcommand{\Snw}{\varphi} 

\newcommand{\dSn}{\Theta} 
\newcommand{\dSnw}{\vartheta} 

\newcommand{\nSn}{\wt{\Sn}}

\newcommand{\ndSn}{\wt{\dSn}}

\newcommand{\mA}{\left< A \right>} 
\newcommand{\mSn}{\left< \Sn \right>} 
\newcommand{\rSn}{\Sn_{\text{rms}}} 
\newcommand{\mdSn}{\left< \dSn \right>} 
\newcommand{\rdSn}{\dSn_{\text{rms}}} 

\newcommand{\Snm}{\Sn_{-}}
\newcommand{\rms}{\text{rms}}
\newcommand{\rmd}{\text{d}} 
\newcommand{\td}{\tau_{\text{d}}} 
\newcommand{\tw}{\tau_{\text{w}}} 
\newcommand{\eX}{X} 
\newcommand{\edT}{{\Delta T}} 
\newcommand{\avT}{\left<\Delta T\right>} 
\newcommand{\tavT}{\langle \Delta T\rangle} 

\DeclareMathOperator\erf{erf}
\DeclareMathOperator\erfc{erfc}
\DeclareMathOperator\erfi{erfi}
\DeclareMathOperator\ei{Ei}

\begin{document}

\title{Level crossings and excess times due to a super-position of uncorrelated exponential pulses}

\author{A.~Theodorsen}
\email{audun.theodorsen@uit.no}

\author{O.~E.~Garcia}
\email{odd.erik.garcia@uit.no}

\affiliation{Department of Physics and Technology, UiT The Arctic University of Norway, N-9037 Troms{\o}, Norway}

\date{\today}

\begin{abstract}
    A well--known stochastic model for intermittent fluctuations in physical systems is investigated. The model is given by a super-position of uncorrelated exponential pulses, and the degree of pulse overlap is interpreted as an intermittency parameter. Expressions for excess time statistics, that is, the rate of level crossings above a given threshold and the average time spent above the threshold, are derived from the joint distribution of the process and its derivative. Limits of both high and low intermittency are investigated and compared to previously known results. In the case of a strongly intermittent process, the distribution of times spent above threshold is obtained analytically. This expression is verified numerically, and the distribution of times above threshold is explored for other intermittency regimes. The numerical results compare favorably to known results for the distribution of times above the mean threshold for an Ornstein--Uhlenbeck process. This contribution generalizes the excess time statistics for the stochastic model which find applications in a wide diversity of natural and technological systems.
\end{abstract}

\maketitle

\section{Introduction}
A stochastic process given by a super-position of uncorrelated pulses can be considered as a reference model for intermittent fluctuations in physical systems. It has found applications in a broad range of fields, including economics, electronics, fission chambers, magnetically confined fusion plasmas, meteorology, oceanography and optics.\cite{campbell-1909, rice-1944, segal-1985, fesce-1986, jang-2004,claps-2005, lefebvre-2008, garcia-prl-2012, elter-2015, dalmao-2015} In many of these applications, the failure or survival of the system depends sensitively on the frequency of large-amplitude fluctuations and the duration of times spent above a critical threshold level. Accordingly, much work has been done in order to calculate the rate of level crossings and average excess times above a threshold level.\cite{rice-1945,bar-david-1972, barakat-1988, leadbetter-2004, alili-2005, daly-2010, yi-2010, bierme-2012, theodorsen-pop-2016, yura-2010}

This contribution is primarily motivated by turbulent flows in the boundary region of magnetically confined plasmas. Evidence points towards these fluctuations being caused by filamentary structures transporting particles and heat towards main chamber walls.\cite{dippolito-2011,zweben-2007} Experimental results provide strong evidence that large-amplitude plasma fluctuations in the boundary region can be described as a super-position of uncorrelated pulses with fixed, exponential pulse shape of constant duration and exponentially distributed pulse amplitudes, with exponentially distributed waiting times between the pulse arrivals.\cite{garcia-pop-2013,garcia-nf-2015,theodorsen-ppcf-2016,kube-2016,theodorsen-ppcf-2016,garcia-nme-2016,theodorsen-nf-gpi,kube-cmod} A stochastic model with these properties has Gamma distributed amplitudes, a parabolic relation between the skewness and flatness moments, an exponential autocorrelation function and a Lorenzian power spectrum.\cite{garcia-prl-2012,garcia-pop-2016,theodorsen-ps-2017}

This stochastic model can be extended in several ways, including adding a noise term,\cite{theodorsen-ps-2017} using different pulse shapes\cite{lowen-teich-fractal,pecseli-fps,garcia-pop-2017-1,garcia-pop-2017-2} or distributions of amplitudes,\cite{daly-2010,theodorsen-ppcf-2016} or allowing for a distribution of pulse durations.\cite{garcia-pop-2017-1,garcia-pop-2017-2} In this contribution, which is an extended version of \Ref{theodorsen-pop-2016}, the rate of threshold crossings and average time above a given threshold are derived and discussed in the case of exponential pulses with fixed duration and shape.

Given the joint probability density function (PDF) $P_{\Sn \dot{\Sn}}(\Sn,\dot{\Sn})$ for a stationary random variable $\Sn(t)$ and its derivative $\dot{\Sn}=\rmd\Sn/\rmd t$, the number of up-crossings of the level $\Phi$ in a time interval of duration $T$ is given by integrating over all positive values of the derivative,\cite{rice-1945,kristensen-1991,sato-2012,fattorini-2012}
\begin{equation}\label{rice}
\eX(\Sn) = T \int\limits_{0}^{\infty} \rmd\dot{\Sn}\, \dot{\Sn} P_{\Sn \dot{\Sn}} ( \Sn,\dot{\Sn} ) .
\end{equation}
For independent, normally distributed $\Sn$ and $\dot{\Sn}$, this gives the celebrated result known as the Rice formula,\cite{rice-1945,kristensen-1991,sato-2012,fattorini-2012,dalmao-2015}
\begin{equation} \label{ricenorm}
\eX(\Sn) = T\, \frac{\dot{\Sn}_\rms}{2\pi\Sn_\rms} \exp{\left( - \frac{\left( \Sn-\mSn \right)^2}{2 \rSn^2} \right)} ,
\end{equation}
where $\tmean{\Sn}$ is the mean value of $\Sn$ and $\Sn_\rms$ and $\dot{\Sn}_\rms$ are the standard deviation or root mean square (rms) values of $\Sn$ and $\dot{\Sn}$, respectively. Here and in the following, $\tmean{\bullet}$ denotes an average over all random variables. The number of level crossings is clearly largest for threshold values close to the mean value of $\Sn$. In this contribution, we will frequently use the normalization
\begin{equation}
    \nSn = \frac{\Sn-\mSn}{\rSn},
    \label{eq:norm-def}
\end{equation}
giving
\begin{equation} \label{ricenorm-normed}
\eX(\nSn) = T\, \frac{\dot{\Sn}_\rms}{2\pi\Sn_\rms} \exp{\left( - \frac{\nSn^2}{2} \right)} .
\end{equation}

The average time $\tavT$ spent above a threshold value $\Sn$ by the stationary process is given by the ratio of the total time spent above the level $\Sn$ to the number of up-crossings $\eX$ in an interval of duration $T$. The former is by definition given by $T [ 1-C_\Sn(\Sn)]$, where $C_\Sn(\Sn)$ is the cumulative distribution function (CDF) of $\Sn$. This gives the average excess time as
\begin{equation}
    \avT\left( \Sn \right) =\frac{T\left[1-C_\Sn(\Sn)\right]}{\eX\left( \Sn \right)} .
\end{equation}
For jointly normally distributed $\Sn$ and $\dot{\Sn}$ with zero correlation (that is, the processes are independent), the average excess time is given by \cite{kristensen-1991,sato-2012,fattorini-2012}
\begin{equation}
    \label{eq:rice-avT}
    \avT(\nSn) = \pi\,  \frac{\rSn}{\dot{\Sn}_\rms}\erfc\left( \frac{\nSn}{\sqrt{2}} \right)\exp\left( \frac{\nSn^2}{2 } \right) ,
\end{equation}
where $\erfc(x)$ denotes the complementary error function (and $\erf(x)$ is the error function).\cite{nist-dlmf} Here and in the following, $x$ denotes a real, unitless variable, used in definitions of special functions. It should be noted that the standard deviation of $\dot{\Sn}$, which appears in both \Eqsref{ricenorm-normed} and \eqref{eq:rice-avT}, is challenging to estimate from measurement data, thus limiting the usefulness of the expressions above. 

The goal of this contribution is to derive expressions for level crossing rates and excess times for a filtered Poisson process (FPP). In \Secref{sec:FPP}, the FPP with a two-sided exponential pulse shape with fixed pulse duration time and exponentially distributed amplitudes is introduced, and some of its statistical properties are reviewed. The derivative of the process is discussed and the joint PDF between the process and its derivative is derived. In \Secref{sec:excess-stat}, expressions for the rate of level crossings and the average excess time for the FPP are given. Limits of a one-sided exponential pulse shape, the normal limit and the limit of strong intermittency are discussed in detail. In \Secref{sec:PDF-edT-an}, we discuss the distribution of excess times in the strong intermittency limit and in the normal limit. \Secref{sec:monte-carlo} gives numerical results for the distribution of excess times in the general case, and compares this to the analytic expressions from \Secref{sec:PDF-edT-an}. The convergence of the rate of level crossings to its analytic expression is also considered in \Secref{sec:PDF-edT-an}. Concluding remarks are given in \Secref{sec:conclusion}. 

\section{The filtered Poisson process}\label{sec:FPP}
In this section, the FPP is introduced and its general features are discussed. First, we present the distribution and moments of the FPP. Secondly, the derivative of the FPP is derived, and its distribution and moments are presented. Lastly, we derive and discuss the joint PDF between the FPP and its derivative.

\subsection{Super-position of pulses}
The FPP can be described as a super-position of uncorrelated pulses, \cite{rice-1944,garcia-prl-2012,garcia-pop-2016,kube-2015,kube-2016,theodorsen-pop-2016,theodorsen-ppcf-2016,garcia-nme-2016,theodorsen-nf-gpi,kube-cmod,campbell-1909,parzen-sp,lowen-teich-fractal,pecseli-fps}
\begin{equation} \label{PhiK}
    \Sn_K(t) = \sum_{k=1}^{K(T)} A_k \Snw\left( \frac{t-t_k}{\td} \right) ,
\end{equation}
where for event $k$, $t_k$ is the pulse arrival time and $A_k$ is the pulse amplitude. The pulse duration time $\td$ and the pulse shape $\Snw(x)$ are assumed to be the same for all events. We will assume the waiting time between pulses to be uncorrelated and exponentially distributed with mean waiting time $\tw$. From this it follows that $K(T)$ is Poisson distributed with constant rate $1/\tw$,
\begin{equation}
P_K(K) = \frac{1}{K!} \left( \frac{T}{\tw} \right)^K \exp\left( -\frac{T}{\tw} \right),
\end{equation}
and therefore that the pulse arrival times $t_k$ are uniformly distributed on $[0,T]$.

\begin{figure}
  \centering
  \includegraphics[width = \textwidth]{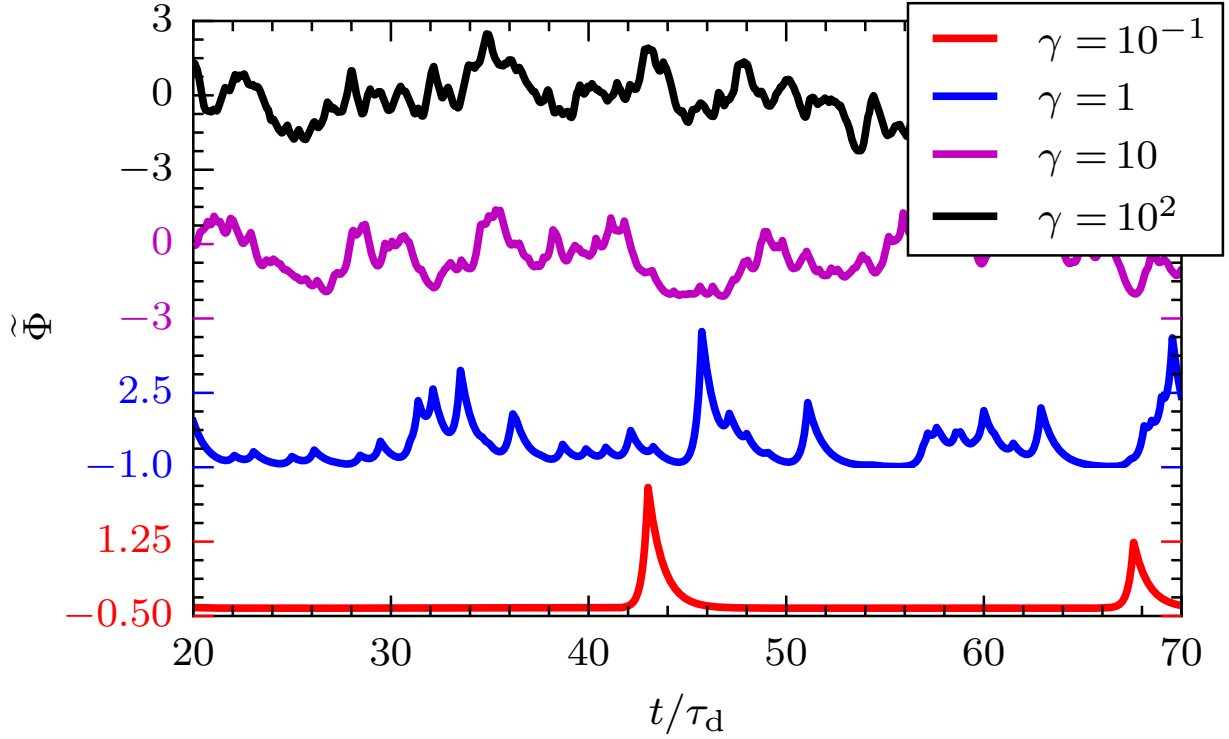}
  \caption{Realizations of the stochastic process for pulse asymmetry parameter $\lambda = 1/4$ and various values of the intermittency parameter $\gamma$.}
  \label{fig:realizations}
\end{figure}

In the following, the pulse shape is described by a two-sided exponential function
\begin{equation}\label{eq:pulse}
\Snw(x) = 
\begin{cases}
\exp{\left(  x / \lambda \right)}, & \quad x<0 ,
\\
\exp{\left( - x/(1-\lambda) \right)}, & \quad x \geq 0 ,
\end{cases}
\end{equation}
where $\lambda$ is a pulse asymmetry parameter restricted to the range $0<\lambda<1$. The ratio between the pulse duration and average waiting time,
\begin{equation}
\gamma = \frac{\td}{\tw} ,
\label{eq:gamma-def}
\end{equation}
is called the \emph{intermittency parameter}, and determines the degree of pulse overlap. It is the most fundamental parameter of the stochastic model. Realizations of this process for various values of $\gamma$ are shown in \Figref{fig:realizations}, using the normalization given in \Eqref{eq:norm-def}. For small $\gamma$, the pulses are well separated and the process is strongly intermittent. For large $\gamma$, there is significant pulse overlap and realizations of the process resembles random noise, with relatively small and symmetric fluctuations around the mean value. For intermediate $\gamma$, large-amplitude bursts can be constructed from one separate large-amplitude pulse, or several smaller amplitude pulses. Because of this, the parameter $\gamma$ can be interpreted as an intermittency parameter for the process, with low values of $\gamma$ giving a highly intermittent process and high values of $\gamma$ giving a weakly intermittent process.

Assuming the PDF of the pulse amplitudes $A$ is an exponential distribution,
\begin{equation}
    P_A(A) = \frac{1}{\mA} \exp\left( - \frac{A}{\mA} \right),\,A>0,
\end{equation}
the stationary distribution of the random variable $\Phi(t)$ can be shown to be a Gamma distribution with shape parameter $\gamma$ and scale parameter $\mA$;\cite{bondesson-1982,daly-2010,garcia-prl-2012,garcia-pop-2016,garcia-pop-2017-2}
\begin{equation}
  P_\Sn(\Sn) = \frac{1}{\mA \Gamma(\gamma)} \left( \frac{\Sn}{\mA} \right)^{\gamma-1} \exp\left( -\frac{\Phi}{\mA} \right),\,\Sn>0,
  \label{eq:pdf_shotnoise}
\end{equation}
where $\Gamma(x)$ is the gamma function.\cite{nist-dlmf} The complementary CDF of $\Sn$ is then given by
\begin{equation}
  1-C_\Sn(\Sn) = Q\left( \gamma , \gamma \Sn/\mSn \right),
  \label{eq:sn_ccdf}
\end{equation}
where $Q(a,x)$ is the regularized upper incomplete gamma function with parameter $a$.\cite{nist-dlmf} In this contribution, we will also use the upper incomplete gamma function $\Gamma(a,x)$.\cite{nist-dlmf} $Q(a,x)$ is defined as $Q(a,x)=\Gamma(a,x)/\Gamma(a)$.

\begin{figure}
  \centering
    \includegraphics[width = \textwidth]{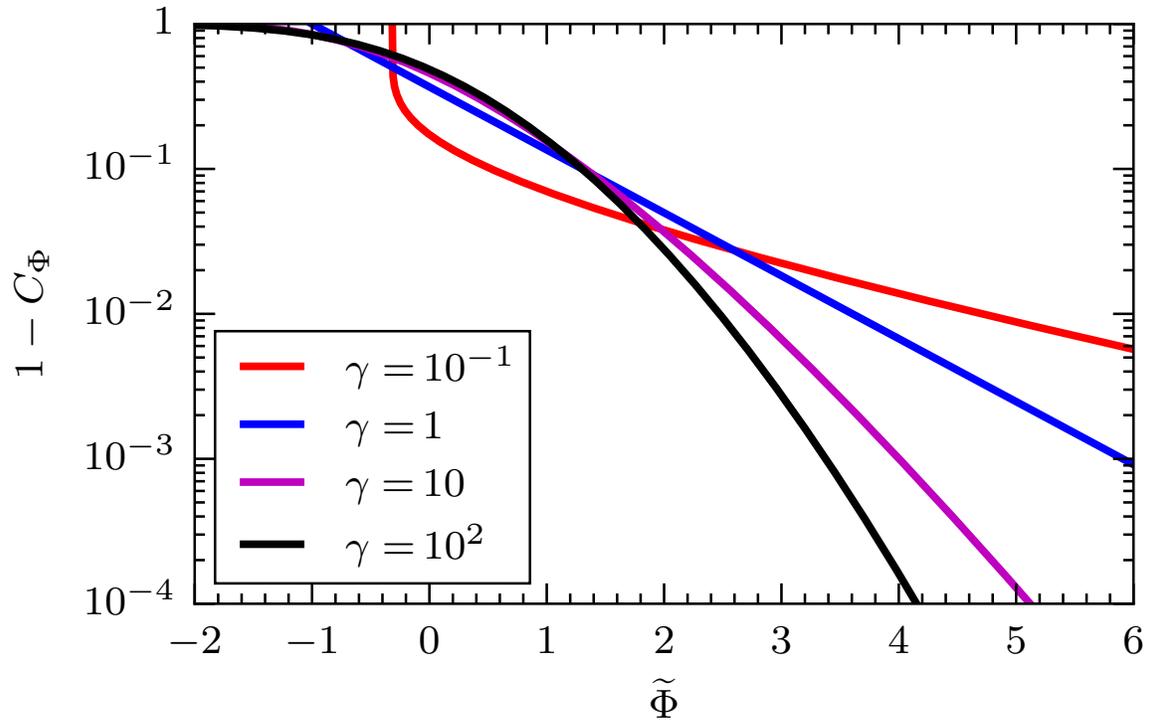}
  \caption{The complementary cumulative distribution function of the  stochastic process for various values of the intermittency parameter $\gamma$.}
  \label{fig:ccdfSn}
\end{figure}

The complementary CDF of $\Sn$ as a function of $\nSn$ for various values of $\gamma$ is presented in \Figref{fig:ccdfSn}. This function can be interpreted as the fraction of time a signal spends above the threshold $\nSn$. As $\gamma$ increases, the PDF approaches a normal distribution. In the normal regime $\gamma \gg 1$, the fraction of time above threshold falls rapidly with increasing threshold level since the fluctuations in the signal are concentrated around the mean value. In the strong intermittency regime, $\gamma \ll 1$, the signal spends long periods of time close to zero value as few pulses overlap significantly. Thus, the total time above threshold increases rapidly as the threshold approaches zero. Also note that for large values of $\Sn$, the total time above threshold is orders of magnitude higher for a process with high intermittency than for a process with low intermittency. For $\gamma = 1$, the PDF of $\Sn$ is an exponential distribution.
 
Correspondingly, the characteristic function of $\Sn$ is 
\begin{equation}
    \mean{\exp(i \Sn u)} = (1 - i \mA u)^{-\gamma}.
    \label{eq:charfun_sn}
\end{equation}
It can likewise be shown that the cumulants of the process for arbitrary pulse shape and amplitude distribution are given by\cite{garcia-prl-2012,garcia-pop-2016}
\begin{equation}
  \kappa_n = \gamma \moment{A}{n} I_n,
  \label{eq:cumulants-def}
\end{equation}
where
\begin{equation}
  I_n = \int \limits_{-\infty}^{\infty} \rmd x\, \left[ \Snw(x) \right]^n.
  \label{eq:integrals}
\end{equation}
For the pulse shape given in \Eqref{eq:pulse}, $I_n = 1/n$ and using that $\tmean{A^n} = n!\,\mA^n$ for exponentially distributed amplitudes, the cumulants for $\Sn$ are\cite{garcia-prl-2012,garcia-pop-2016}
\begin{equation}
  \kappa_n = (n-1)!\, \gamma \mA^n.
  \label{eq:cumulants-sn}
\end{equation}
Note that the cumulants, and therefore also the PDF, are independent of the pulse asymmetry parameter $\lambda$. Given the cumulants, we can find the lowest order moments of the process:\cite{rice-1944,bondesson-1982,garcia-prl-2012,garcia-pop-2016}
\begin{subequations}
    \label{eq:moments-sn}
\begin{align}
  \mSn &= \gamma \mA,\label{eq:mean-sn} \\
  \rSn^2 &=  \gamma \mA^2, \label{eq:rms-sn}\\
  S_{\Sn} &= \frac{2}{\gamma^{1/2}}, \label{eq:skew-sn}\\
  F_{\Sn} &= 3+ \frac{6}{\gamma}.\label{eq:flat-sn}
\end{align}
\end{subequations}
Here, $S_\Sn$ is the skewness of the random variable $\Sn$, and $F_\Sn$ is its flatness. The relative fluctuation level is $\rSn / \mSn = 1/\gamma^{1/2}$. There is a parabolic relation between skewness and flatness: $F_\Sn\left( S_\Sn \right) = 3+3 S_\Sn^2 / 2$. It can be shown that the distribution of the normalized process $\nSn = (\Sn-\mSn)/\rSn$ resembles a standard normal distribution (that is, a normal distribution with zero mean and unit standard deviation) in the limit $\gamma \to \infty$, independent of pulse shape and amplitude distribution.\cite{rice-1944} In this case, both the skewness $S_{\Sn}$ and the excess kurtosis $F_{\Sn}-3$ vanish.\cite{rice-1944,garcia-prl-2012} Conversely, for $\gamma \to 0$, the skewness and kurtosis moments both tend to infinity. Note that from the definition of skewness and flatness, it follows that $S_\Sn = S_{\nSn}$ and $F_\Sn = F_{\nSn}$.

Note that $\Sn$ is non-negative, giving $\nSn \geq - \gamma^{1/2}$. By contrast, a normally distributed random variable has infinite support. The difference between the PDF of $\nSn$ and a standard normal distribution due to this discrepancy is negligible in practice, since values of $-\gamma^{1/2}$ or less are highly unlikely for a standard normal distribution in the case of $\gamma \gg 1$.

\subsection{The derivative of the filtered Poisson process}\label{sec:derivative-shot-noise}
In order to calculate the joint distribution of the process and its derivative, the normalized time derivative is defined by
\begin{equation}
\dSn_K(t) = \frac{\td}{2} \frac{\text{d}\Sn_K}{\text{d} t} 
= \sum_{k = 1}^{K(T)} A_k \dSnw{\left( \frac{t-t_k}{\td} \right)} ,
\end{equation}
where the pulse shape is given by
\begin{equation}
    \dSnw(s) = \frac{1}{2} \frac{\rmd \Snw}{\rmd s} = \frac{1}{2}
\begin{cases}
(\lambda)^{-1} \exp{\left(  s/ \lambda \right)}, & \quad s<0 ,
\\
-(1-\lambda)^{-1} \exp{\left( - s/(1-\lambda) \right)}, & \quad s \geq 0 .
\\
\end{cases}
\end{equation}
Here, we have divided by a factor 2 in order for the pulse shape to fulfill $\int_{-\infty}^\infty \rmd s \, \abs{\dSnw(s)}=1$.\footnote{This is a correction to $\dSnw$ given in \Ref{theodorsen-pop-2016}. This leads to corrections in the rms-value of $\dSn$ and the joint PDF between $\Sn$ and $\dSn$, but not the excess statistics.} This is another stochastic process of the same type as that given in \Eqref{PhiK}, but with a different pulse shape. Since the process $\Sn(t)$ is stationary, it follows that $\mean{\dSn}=0$, which is easily verified from \Eqref{eq:cumulants-sn}. The processes $\Sn(t)$ and $\dSn(t)$ are evidently dependent yet also uncorrelated,
\begin{equation}
\mean{\Sn \dSn} = \frac{\td}{4}\frac{\text{d}}{\text{d}t} \mean{\Sn^2} = 0 .
\end{equation}
In \Appref{app:jpdf-normal-limit}, the joint PDF between $\Sn$ and $\dSn$ is used to demonstrate that $\Sn$ and $\dSn$ become independent in the limit $\gamma \to \infty$.

The lowest order moments of $\dSn$ are readily calculated as 
\begin{subequations}
    \label{eq:moments-dsn}
\begin{align}
  \mdSn &= 0,\label{eq:mean-dsn} \\
  \rdSn^2 &=  \gamma \mA^2 / \left[ 4 \lambda (1-\lambda) \right], \label{eq:rms-dsn}\\
  S_{\dSn} &= 2(1-2\lambda)/[\gamma \lambda (1-\lambda)]^{1/2}, \label{eq:skew-dsn}\\
  F_{\dSn} &= 3+6[1+(1-2\lambda)^2/\lambda (1-\lambda)]/\gamma.\label{eq:flat-dsn}
\end{align}
\end{subequations}
In the limit of $\lambda \to 0$ or $\lambda \to 1$, the moments $\rdSn$, $S_\dSn$ and $F_\dSn$ diverge, meaning the PDF of $\dSn$ does not exist in this case. In these limits, the pulse shape in $\Sn$ is discontinuous and the derivative of the pulse shape contains delta functions. Thus we require the two-sided exponential pulse shape to calculate the rate of level crossings. It will later be shown that these limits exist for the rate of level crossings and are consistent with other methods starting from the one-sided exponential pulse shape. Thus, while this method cannot be used to calculate the rate of level crossings for a discontinuous signal, the rate still exists.\cite{bar-david-1972,daly-2010,bierme-2012}

\begin{figure}
  \centering
  \includegraphics[width = \textwidth]{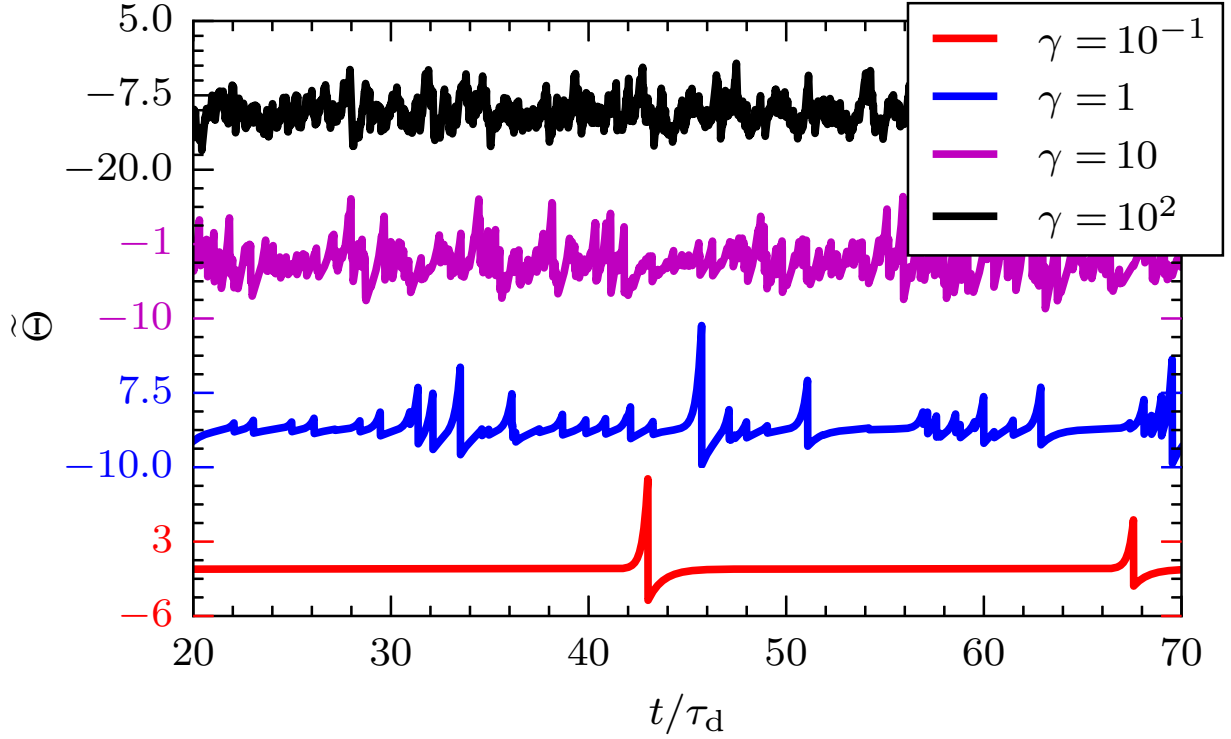}
  \caption{Realizations of the derivative of the stochastic process for asymmetry parameter $\lambda = 1/4$ and various values of the intermittency parameter $\gamma$.}
  \label{fig:dsn_realizations}
\end{figure}

Using the same approach as in \Refs{garcia-prl-2012} and \onlinecite{garcia-pop-2016}, the characteristic function of $\dSn$ is given by
\begin{equation}
    \mean{\exp(i \dSn v)} = \left(1-i \mA \frac{v}{2 \lambda}\right)^{-\lambda \gamma} \left(1 + i \mA \frac{v}{ 2(1-\lambda)}\right)^{-(1-\lambda)\gamma}.
    \label{eq:charfun_dsn}
\end{equation}
This characteristic function can be interpreted as originating from the sum of two independent gamma distributed variables, one over positive values with shape parameter $\gamma \lambda$ and scale parameter $\mA/(2 \lambda)$, and the other over negative values with shape parameter $\gamma (1-\lambda)$ and scale parameter $\mA/[2 (1-\lambda)]$.  The PDF of this compound process is a convolution of the two gamma distributions, which to the best of the authors knowledge does not have a closed form. Still, the argument in \Refs{rice-1944,garcia-prl-2012,garcia-pop-2016} applies here as well, and the PDF of $\dSn$ resembles a normal distribution in the limit $\gamma \to \infty$. In \Figref{fig:dsn_realizations}, realizations for $\ndSn$ are presented for $\lambda = 1/4$ and various values of $\gamma$. Arrival times and pulse amplitudes are the same as in \Figref{fig:realizations}. Again, the process is strongly intermittent for low values of $\gamma$, and resembles random noise for high values of $\gamma$.

\begin{figure}
  \centering
  \includegraphics[width = \textwidth]{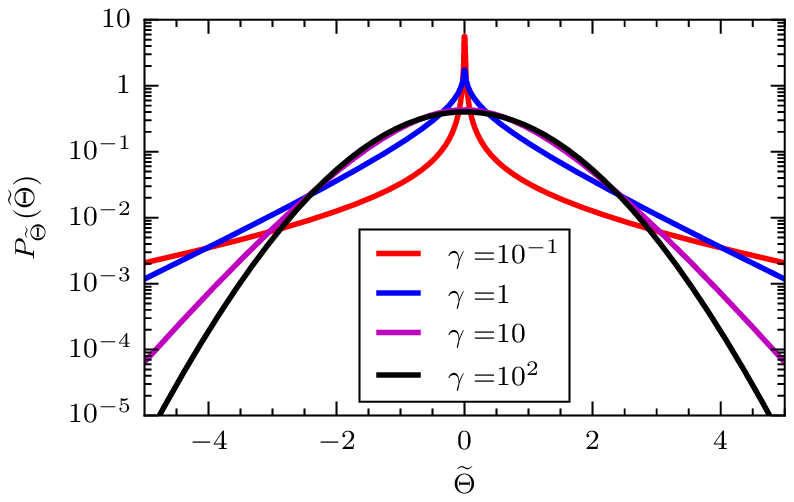}
  \caption{PDF of the normalized derivative of the stochastic process with asymmetry parameter $\lambda = 1/2$ and various values of the intermittency parameter $\gamma$.}
  \label{fig:dsn_pdf}
\end{figure}

\begin{figure}
  \centering
  \includegraphics[width = \textwidth]{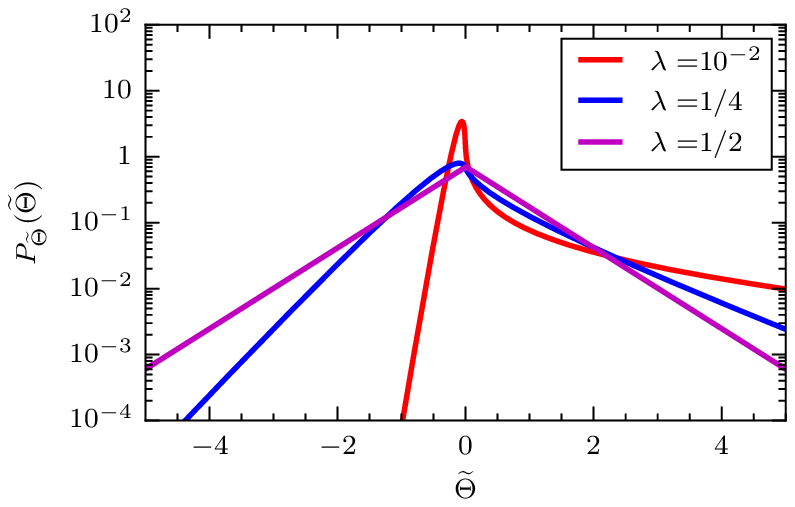}
  \caption{PDF of the normalized derivative of the stochastic process with intermittency parameter $\gamma = 2$ and various values of the asymmetry parameter $\lambda$.}
  \label{fig:dsn_pdf_lam}
\end{figure}

By choosing $\lambda = 1/2$, the pulse shape $\Snw(x)$ is symmetric. In this case, the characteristic function in \Eqref{eq:charfun_dsn} has an inverse transformation in closed form, and the corresponding PDF is given by
\begin{equation}
    \label{eq:pdf-dsn-symmetric}
    P_{\ndSn}(\ndSn) = \sqrt{\frac{2\gamma}{\pi}} \frac{2^{-\gamma/2}}{\Gamma(\gamma/2)} \abs{ \sqrt{\gamma} \ndSn }^{(\gamma-1)/2} \mathcal{K}_{(\gamma-1)/2}\left( \abs{ \sqrt{\gamma} \ndSn} \right),
\end{equation}
where $\mathcal{K}_a(x)$ is the modified Bessel function of the second kind.\cite{nist-dlmf} This PDF is presented in \Figref{fig:dsn_pdf} for various values of $\gamma$. For small values of $\gamma$, this PDF has exponential tails and is sharply peaked at the mean value, while it resembles a normal distribution for large values of $\gamma$. The same PDF for $\gamma = 2$ and various values of $\lambda$ is presented in \Figref{fig:dsn_pdf_lam}. As the asymmetry parameter approaches $0$, the skewness and flatness of $\dSn$ increases. It can be seen from \Eqref{eq:charfun_dsn} that in the case $\lambda = 1/2$ and $\gamma = 2$, $\dSn$ is symmetrically Laplace distributed with zero mean and standard deviation $\dSn_\rms = 2 \mA^2$.\cite{kozubowski-2000}

\subsection{The joint PDF of the filtered Poisson process}\label{sec:jpdf}
The joint PDF of $\Sn$ and $\dSn$ is generally given by 
\begin{equation}
  P_{\Sn \dSn} \left( \Sn, \dSn \right) = \frac{1}{\left( 2 \pi \right)^2} \int\limits_{-\infty}^{\infty} \text{d} u \, \int\limits_{-\infty}^{\infty} \text{d}v \, \exp\left( -i \Sn u -i \dSn v \right) \mean{\exp\left( i u \Sn + i v \dSn \right)}.
  \label{jpdf-start}
\end{equation}
Using that individual events are uncorrelated and that the number of pulses is Poisson distributed, the characteristic function of $\Sn$ and $\dSn$ can be calculated as
\begin{equation}
  \mean{\exp\left( i u \Sn + i v \dSn \right)} = \exp\left( \gamma \int\limits_{-\infty}^{\infty}\text{d}A P_A(A)\, \int\limits_{-\infty}^{\infty} \text{d}s\, \left[  \exp\left( i u A \Snw(s) + i v A \dSnw(s) \right)-1 \right]  \right).
  \label{jpdf-charfun-start}
\end{equation}
This expression is given in \Refs{pecseli-fps,sato-2012} for the case of fixed (degenerately distributed) pulse amplitudes, although the generalization is straightforward. Exchanging the order of integration, we find that 
\begin{equation}
    \mean{\exp(i u \Sn + i v \dSn)} =\left[ 1 - i \mA \left( u + \frac{v}{2 \lambda} \right) \right]^{-\gamma \lambda} \left[ 1- i \mA \left( u-\frac{v}{2(1-\lambda)} \right) \right]^{-\gamma\left( 1-\lambda \right)}.
    \label{eq:towards_joint_pdf}
\end{equation}
We note that we recover the expression for the characteristic function of $\Sn$ in \Eqref{eq:charfun_sn} by setting $v=0$ in this equation, and we recover the characteristic function of $\dSn$ in \Eqref{eq:charfun_dsn} by setting $u=0$.
Substituted into \Eqref{jpdf-start}, the stationary joint PDF can be obtained in closed form. We change variables to $x = \mA [u+v/(2 \lambda)]$ and $y = \mA \{u-v/[2(1-\lambda)]\}$, and use the notation
\begin{align*}
  \alpha &= \frac{\lambda}{\mA}\left[ \Sn + 2(1-\lambda)\dSn \right], \\
  \beta &= \frac{1-\lambda}{\mA} \left( \Sn - 2 \lambda \dSn \right).
\end{align*}
The joint PDF can now be written as
\begin{equation}
  P_{\Sn \dSn}\left( \Sn, \dSn \right) = \frac{2 \lambda (1-\lambda)}{\left( 2 \pi \mA \right)^2} \int\limits_{-\infty}^{\infty} \text{d}x\, \left[ 1-i x \right]^{-\gamma \lambda} \exp\left( -i \alpha x \right)    \int\limits_{-\infty}^{\infty}\text{d}y\, \left[ 1-i y \right]^{-\gamma (1-\lambda)} \exp\left( - i \beta y \right) .
  \label{eq:shot_noise_jpdf_alt_1}
\end{equation}
The integrals can be performed separately, and we get the closed form expression 
\begin{equation} \label{jpdf}
  P_{\Sn \dSn}{\left( \Sn, \dSn \right)} = \frac{2 \gamma^\gamma \lambda^{\gamma \lambda} (1-\lambda)^{\gamma(1-\lambda)}  }{\mSn^\gamma \Gamma(\gamma \lambda) \Gamma{\left( \gamma (1-\lambda) \right)}} \exp\left(-\frac{\gamma\Sn}{\mSn}\right) \left[ \Sn+2(1-\lambda)\dSn \right]^{\gamma \lambda -1} \left( \Sn-2\lambda \dSn \right)^{\gamma (1-\lambda) -1}.
\end{equation}
This is non-zero only for the limited range $-\Sn/[2(1-\lambda)]<\dSn<\Sn/(2\lambda)$, which follows from the fact that the signal $\Sn(t)$ cannot decrease faster than the rate of decay of individual pulse structures, nor increase slower than the rate of growth of individual pulses, since the pulse amplitudes are positive definite. The dependence between $\Sn$ and $\dSn$ is evident from \Eqref{jpdf}, since the joint PDF is not separable into a product of the marginal PDFs. As expected, $P_\Sn(\Sn)$ can be recovered by integrating over $\Theta$. Also note that the expression for the joint PDF diverges in the limits $\lambda \to 0$ and $\lambda \to 1$, as was the case for the moments and PDF of $\dSn$.
As the PDFs of both $\Sn$ and $\dSn$ resemble normal distributions in the limit $\gamma \to \infty$ and they are uncorrelated, the joint PDF for $\Sn$ and $\dSn$ resembles the product of two normal distributions, that is, a joint normal distribution with vanishing correlation coefficient. This is demonstrated explicitly in \Appref{app:jpdf-normal-limit}. Thus, in the normal limit $\gamma\to\infty$, the classical Rice formula given by \Eqref{ricenorm} is recovered.
As in the case of $P_\Sn$, there is a discrepancy between $P_{\Sn \dSn}$ and a joint normal distribution due to the limited region of non-zero values of $P_{\Sn \dSn}$. The domain of non-zero values can be written as $-(\nSn+\gamma^{1/2})/(1-\lambda) <\ndSn /\sqrt{\lambda (1-\lambda)}< (\nSn+\gamma^{1/2})/\lambda$, where $\ndSn = \dSn/\dSn_\rms$. For standard normally distributed variables, values outside of this domain are highly unlikely in the case of $\gamma \gg 1$, and this discrepancy is in practice negligible.

\begin{figure}
  \centering
  \begin{subfigure}{0.47\textwidth}
    \includegraphics[width = \textwidth]{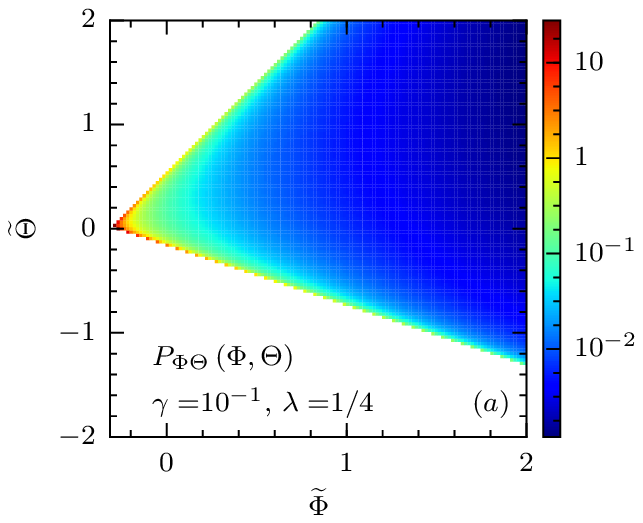}
    \caption{\label{fig:jpdf-g0.1-l0.25}}
  \end{subfigure}
  ~
  \begin{subfigure}{0.47\textwidth}
    \includegraphics[width = \textwidth]{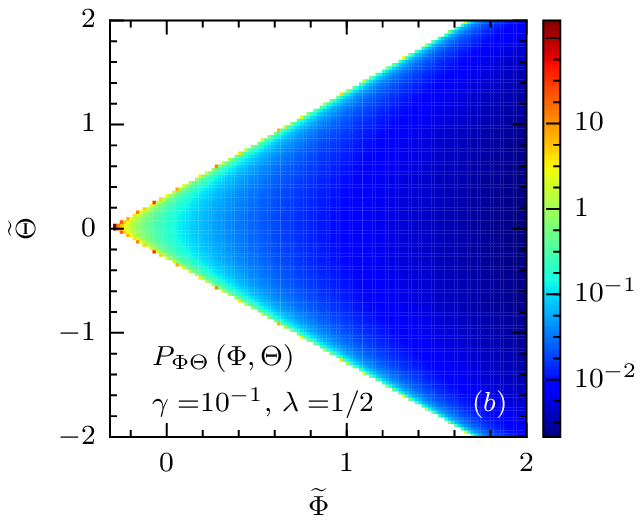}
    \caption{\label{fig:jpdf-g0.1-l0.5}}
  \end{subfigure}
  
  \begin{subfigure}{0.47\textwidth}
    \includegraphics[width = \textwidth]{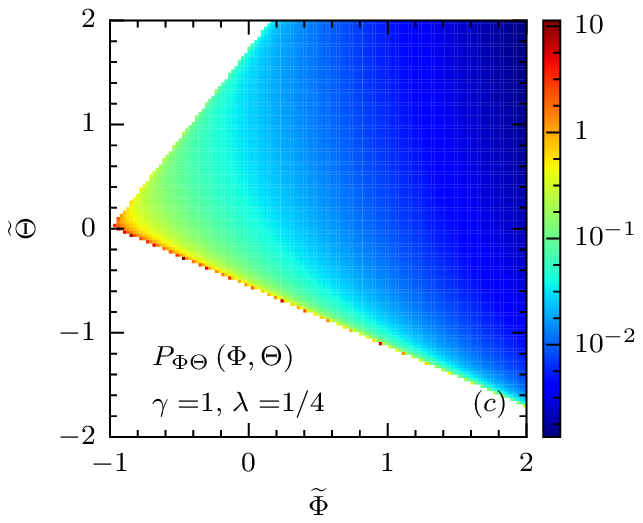}
    \caption{\label{fig:jpdf-g1.0-l0.25}}
  \end{subfigure}
  ~
  \begin{subfigure}{0.47\textwidth}
    \includegraphics[width = \textwidth]{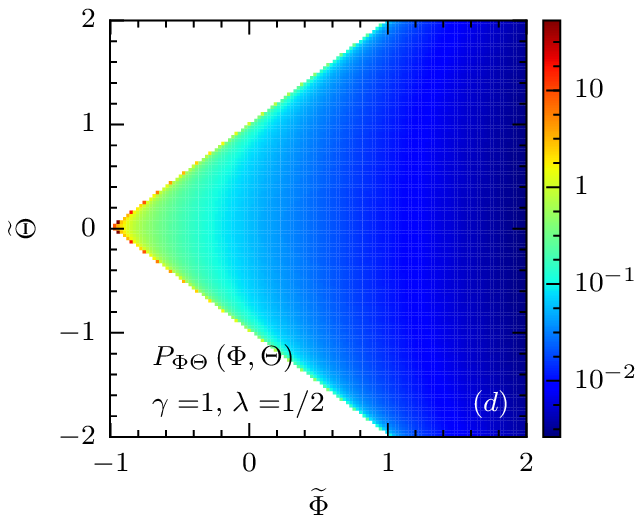}
    \caption{\label{fig:jpdf-g1.0-l0.5}}
  \end{subfigure}

  \begin{subfigure}{0.47\textwidth}
    \includegraphics[width = \textwidth]{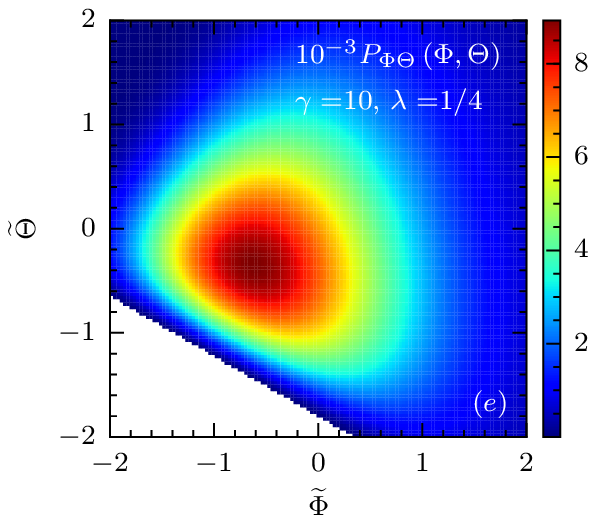}
    \caption{\label{fig:jpdf-g10-l0.25}}
  \end{subfigure}
  ~
  \begin{subfigure}{0.47\textwidth}
    \includegraphics[width = \textwidth]{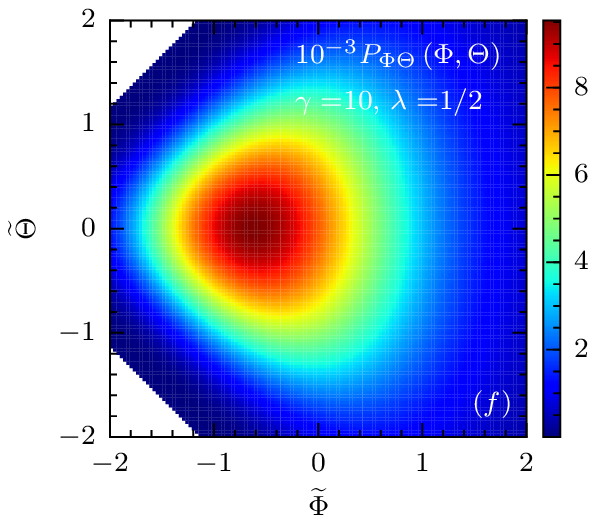}
    \caption{\label{fig:jpdf-g10.0-l0.5}}
  \end{subfigure}

  \caption{\label{fig:jpdf} Joint PDF between $\nSn$ and $\ndSn$ for various values of the pulse asymmetry parameter $\lambda$ and the intermittency parameter $\gamma$.\footnote{For the case denoted $\gamma = 1$, $\lambda = 1/2$, the numerical value $\gamma = 1.01$ is used to avoid singularities in the numerical computation.}}
\end{figure}
The joint distribution $P_{\Sn\dSn}(\Sn,\dSn)$ is presented in \Figref{fig:jpdf} for $\gamma \in \{10^{-1},1,10\}$ and $\lambda \in \{1/4,1/2\}$. It should be noted that logarithmic scaling is used for $\gamma = 10^{-1}$ and $1$, while linear scaling is used for $\gamma = 10$. The white area in all figures are the regions where $P_{\Sn\dSn}$ vanishes, as given by \Eqref{jpdf}. The joint distribution for $\gamma \leq 1$ diverges at $\Sn = 0$ and $\dSn = 0$, corresponding to $\ndSn = 0$, $\nSn = -\gamma^{1/2}$, since the pulses arrive rarely enough for the signal to fall close to zero value for long time durations. In this case, the signals are very likely to decay or grow undisturbed at the rate of individual pulses, explaining the increased value of the joint distribution near the lines $\dSn = -\Sn/[2(1-\lambda)]$, $\dSn = \Sn/(2\lambda)$.

\section{Excess time statistics}\label{sec:excess-stat}
In this section we present the rate of threshold crossings and average time above threshold for the FPP. Limits of one-sided exponential pulse shape, and weak and strong intermittency are explored and compared to previous works.

\subsection{Formulation of excess time statistics}
The rate of up-crossings above a threshold level $\Sn$ is now readily calculated from \Eqsref{rice} and \eqref{jpdf} as
\begin{equation}
\frac{\td}{T} \eX(\Sn) = 2 \int_{0}^{\infty}\text{d}\dSn\, \dSn P_{\Sn \dSn}(\Phi, \dSn) = \frac{\lambda^{\gamma \lambda -1} \left( 1-\lambda \right)^{\gamma\left( 1-\lambda \right)-1}}{\gamma \Gamma\left( \gamma \lambda \right) \Gamma\left( \gamma \left( 1-\lambda \right) \right)} \left( \frac{\gamma \Phi}{\mSn} \right)^\gamma \exp\left( - \frac{\gamma\Phi}{\mSn} \right),
\label{eq:sn_X}
\end{equation}
which, together with the complementary CDF in \Eqref{eq:sn_ccdf}, gives the average time above the threshold for each threshold crossing,
\begin{equation}
\frac{1}{\td}\,\avT(\Phi) = \frac{\gamma \Gamma\left(\gamma\lambda\right)\Gamma\left(\gamma(1-\lambda)\right)}{\lambda^{\gamma \lambda-1} (1-\lambda)^{\gamma (1-\lambda)-1}} Q\left( \gamma, \frac{\gamma \Phi}{\mSn} \right) \left( \frac{\gamma \Phi}{\mSn} \right)^{-\gamma}\exp\left( \frac{\gamma \Phi}{\mSn} \right).
\label{eq:sn_avT}
\end{equation}
Note that both \Eqsref{eq:sn_X} and \eqref{eq:sn_avT} can be written as a pre-factor depending on $\gamma$ and $\lambda$, multiplied by a function of $\gamma$ and the variable 
\begin{equation}
    \label{eq:norm-sn-alt}
\gamma \Sn / \mSn = \sqrt{\gamma} \nSn + \gamma.
\end{equation}
This indicates that the functional shape of both equations with threshold level depend only on the intermittency parameter $\gamma$, while the function value depends on both $\gamma$ and $\lambda$. By contrast, the complementary CDF given by \Eqref{eq:sn_ccdf} does not depend on $\lambda$.

From the joint PDF in \Eqref{jpdf}, it is clear that the dependency between $\Sn$ and $\dSn$ is important for the rate of threshold crossings. In order to investigate the effect of this dependency, we calculate the rate of threshold crossings divided by the PDF of $\Sn$:
\begin{equation}\label{eq:X_div_PSn}
    \frac{\td}{T}\frac{X(\Sn)}{P_\Sn(\Sn)} = \frac{\lambda^{\gamma \lambda} (1-\lambda)^{\gamma(1-\lambda)} \Gamma(1+\gamma)}{\Gamma[1+\gamma(1-\lambda)]\Gamma(1+\gamma \lambda)} \Sn.
\end{equation}
On the other hand, starting from \Eqref{eq:sn_X} and assuming $\Sn$ and $\dSn$ are independent gives
\begin{equation}\label{eq:X_div_PSn_independent}
    \frac{\td}{T}\frac{X(\Sn)}{P_\Sn(\Sn)} = \frac{2 P_\Sn(\Sn) \int_0^\infty \rmd \dSn\, \dSn P_\dSn(\dSn)}{P_\Sn(\Sn)} = 2 \int\limits_0^\infty \rmd \dSn\, \dSn P_\dSn(\dSn),
\end{equation}
which is independent of $\Sn$. Thus an assumption of independence will always give the wrong algebraic factor, although this is not very relevant for large $\Sn$ where the exponential term dominates. Also note that \Eqref{eq:X_div_PSn_independent} gives the correct result in the limit $\gamma \to \infty$, where the process and its derivative are indeed independent. However, inserting the PDF of $\dSn$ from \Secref{sec:derivative-shot-noise} into \Eqref{eq:X_div_PSn_independent} gives a surprisingly complicated result, presented in \Appref{app:half-mean-dsn}. There is significant discrepancy between this expression and the prefactor in \Eqref{eq:X_div_PSn}. Thus accurately accounting for the dependency between $\Sn$ and $\dSn$ is necessary for correctly predicting the rate of threshold crossings.

\begin{figure}
  \centering
    \includegraphics[width = \textwidth]{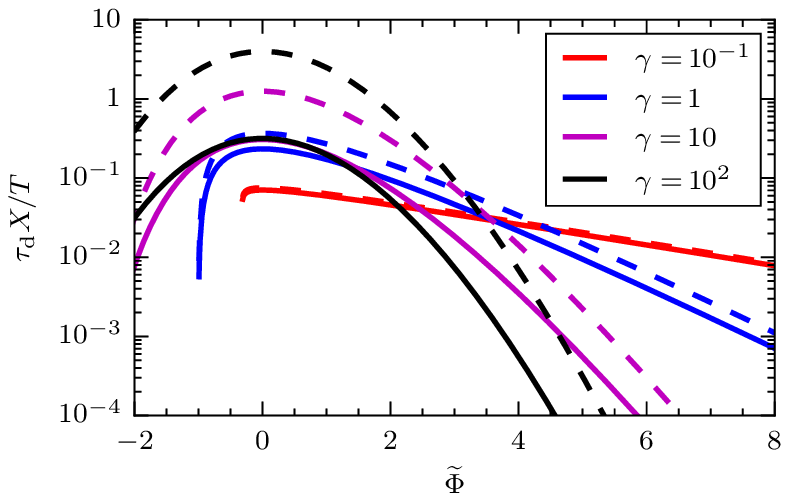}
  \caption{The rate of up-crossings for the stochastic process with pulse asymmetry parameters $\lambda = 1/2$ (full lines) and $\lambda = 0$ (dashed lines) and various values of the intermittency parameter $\gamma$.}
  \label{fig:ex-X}
\end{figure}

The rate of up-crossings as function of the threshold level for various values of $\gamma$ is presented in \Figref{fig:ex-X}. Full lines show the case of $\lambda = 1/2$, while dashed lines show the rate of level crossings in the limit $\lambda \to 0$. The analytical expression in this limit will be discussed further in \Secref{sec:limit-one-sided-waveform}. The total number of crossings is evidently proportional to the length of the time series $T$ and inversely proportional to the pulse duration $\td$. The rate of threshold crossings is highest for thresholds close to the mean value of the process in all cases. In the normal regime $\gamma\gg1$, there are comparatively few crossings for threshold levels much smaller or much larger than the mean value due to the low probability of large-amplitude fluctuations. The rate of level crossings is therefore a narrow Gaussian function in this limit. In the strong intermittency regime, $\gamma\ll1$, the signal spends most of the time close to zero value, and virtually any pulse arrival will give rise to a level crossing for finite threshold values. As seen in \Figref{fig:ex-X}, the rate of level crossings approaches a step function in this limit. For $\lambda = 1/2$, the rate of level crossings at the mean value, $\nSn = 0$, approaches a definite value. In \Secref{sec:ex-norm-lim} this value is shown to be $1/\pi$. In contrast, there is no limiting value for $\lambda = 0$. In this case $X(\nSn = 0) \to \infty$ as $\gamma \to \infty$, as will be demonstrated in \Secref{sec:ex-norm-lim}. 

\begin{figure}
  \centering
    \includegraphics[width = \textwidth]{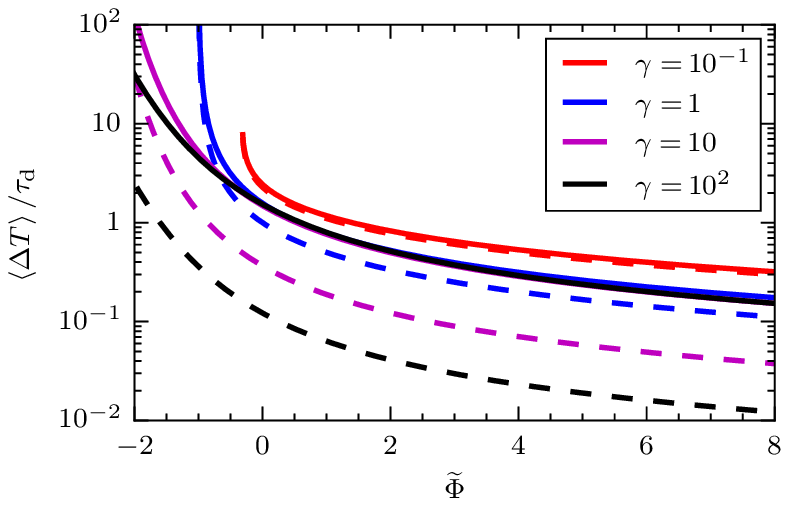}
    \caption{The average time above threshold for the stochastic process with pulse asymmetry parameters $\lambda = 1/2$ (full lines) and $\lambda = 0$ (dashed lines) and various values of the intermittency parameter $\gamma$.}

  \label{fig:ex-avT}
\end{figure}

The average time above threshold is presented in \Figref{fig:ex-avT} for various values of $\gamma$. Full lines show the case of $\lambda = 1/2$, while dashed lines show the average time above threshold in the limit $\lambda \to 0$. While both the rate of threshold crossings and the fraction of time above threshold vary qualitatively as $\gamma$ changes, the shape of  the average time above threshold is fairly similar. In all cases the average excess time decreases monotonically with the threshold level, with a fast drop for small threshold values. This is followed by a slow tapering off for large threshold values. For the range of intermittency parameters considered here, the average excess time is of the order of the pulse duration or shorter for large threshold values. For $\lambda \to 0$, the average time above threshold decreases by about half a decade for each tenfold increase in $\gamma$, but the functional shape varies little. For $\lambda = 1/2$, the average time above threshold converges to the Rice result, as will be shown in \Secref{sec:ex-norm-lim}. It can be shown that for given $\gamma$ and $\lambda$, $\avT /\td$ scales as $1 / \nSn$ in the limit $\nSn \to \infty$.
As the threshold value increases above the mean signal value, up-crossings of the threshold become fewer while the signal spends less time in total above the threshold. Evidently these two effects nearly cancel, and the average excess time decreases slowly with increasing threshold level.

\subsection{Limit of the one-sided pulse shape}\label{sec:limit-one-sided-waveform}
As stated in \Secref{sec:jpdf}, the limit of the one-sided exponential pulse shape does not exist for $P_\dSn$ or $P_{\Sn \dSn}$. This is due to the fact that the pulse shape $\Snw(x)$ is discontinuous in this case, and therefore second and higher order moments of its derivative do not exist. However, the rate of level crossings for the discontinuous process still exists, and has been discussed in for example \Refs{bar-david-1972,bierme-2012,daly-2010}. Taking either of the limits $\lambda \to 0$ and $\lambda \to 1$ give the same result, and yield 
\begin{equation}
    \frac{\td}{T}\eX(\Sn) = \frac{1}{\Gamma(\gamma)} \left(\frac{\gamma \Sn}{\mSn}\right)^{\gamma} \exp\left( - \frac{\gamma \Sn}{\mSn}\right).
    \label{eq:ex-l0}
\end{equation}
This result was also obtained in \Ref{bierme-2012} by considering the Fourier transform of the number of level crossings. Since the complementary CDF of $\Sn$ does not depend on $\lambda$, the total time the signal spends above threshold remains unchanged, and the average time above threshold is simply 
\begin{equation}
    \label{eq:avT-l0}
    \frac{1}{\td} \avT(\Sn) = \Gamma(\gamma) Q\left(\gamma,\frac{\gamma \Sn}{\mSn}\right) \left(\frac{\gamma \Sn}{\mSn}\right)^{-\gamma} \exp\left(\frac{\gamma \Sn}{\mSn}\right).
\end{equation}
The functional shape of \Eqsref{eq:ex-l0} and \eqref{eq:avT-l0} are the same as in the more general expressions given by \Eqsref{eq:sn_X} and \eqref{eq:sn_avT}, since $\lambda$ only appears in the prefactor of these equations.
The approach discussed in \Ref{daly-2010} also leads to the results presented in this section, although they are not explicitly given in the reference.

\subsection{The normal limit}\label{sec:ex-norm-lim}
In the limit of large $\gamma$, the expression for $\eX(\Sn)$ can be simplified and shown to be equal to the case for a normally distributed process. Using Stirling's approximation for the Gamma functions in \Eqref{eq:sn_X}, we have in the normal limit:
\begin{equation}
    \label{eq:stirling-two-gamma}
   \lim_{\gamma \to \infty} \Gamma(\gamma \lambda) \Gamma(\gamma(1-\lambda)) =\lim_{\gamma \to \infty}  2 \pi \gamma^{\gamma-1} \lambda^{\gamma \lambda - 1/2} (1-\lambda)^{\gamma (1-\lambda) -1/2} \exp(-\gamma).
\end{equation}
Inserting this result into \Eqref{eq:sn_X}, and using the normalized threshold in \Eqref{eq:norm-sn-alt}, the rate of crossings in the weak intermittency case $\gamma \gg 1$ can be written as
\begin{equation}
  \lim_{\gamma \to \infty}  \frac{\td}{T}\eX(\nSn) =\lim_{\gamma \to \infty}   \frac{1}{2 \pi \sqrt{\lambda (1-\lambda)}} \left( \frac{\nSn}{\gamma^{1/2}}+1 \right)^{\gamma} \exp\left( -\gamma^{1/2} \nSn \right).
  \label{eq:eX-ginf-start}
\end{equation}
In \Appref{app:normal-limit-1d}, we show that
\begin{equation}
    \label{useful-limit}
    \lim_{\gamma \to \infty} \left( \nSn / \gamma^{1/2} +1 \right)^\gamma \exp\left( - \gamma^{1/2} \nSn \right) = \exp\left(- \nSn^2 /2 \right),
\end{equation}
and the rate of level crossings in the limit $\gamma \to \infty$ can be written as
\begin{equation}
    \lim_{\gamma \to \infty} \frac{\td}{T}\eX(\nSn) =  \frac{1}{2 \pi \sqrt{\lambda (1-\lambda)}} \exp\left(-\nSn^2/2\right).
  \label{eq:eX-ginf}
\end{equation}
This expression is equal to \Eqref{ricenorm-normed}, when using $\rSn$ from \Eqref{eq:rms-sn} and $\dot{\Sn}_\rms = 2 \rdSn/ \td$ from \Eqref{eq:rms-dsn}.  As mentioned in the discussion of \Figref{fig:ex-X}, in the case of $\lambda = 1/2$, we have that $\lim_{\gamma \to \infty} \td X(\nSn = 0) / T = 1/\pi$.

In \Appref{app:Q-to-erfc}, it is shown that
\begin{equation}
    \label{eq:Q-to-erfc}
    \lim_{\gamma \to \infty} Q\left(\gamma,\sqrt{\gamma} \nSn +\gamma\right) = \frac{1}{2} \erfc\left(\frac{\nSn}{\sqrt{2}}\right),
\end{equation}
and the expression for the average time above threshold in \Eqref{eq:sn_avT} can be shown to be equivalent to the expression given by \Eqref{eq:rice-avT} in the case $\gamma \to \infty$. Note that for $\lambda = 1/2$, we have the limit $\lim_{\gamma \to \infty} \avT(\nSn = 0)/\td = \pi/2$.

Starting from \Eqref{eq:ex-l0} and going through the same procedure as above, we have in the cases $\lambda =0$ and $\lambda = 1$
\begin{equation}
    \lim_{\gamma \to \infty} \frac{\td}{T}\frac{\eX(\nSn)}{\sqrt{\gamma}} =  \frac{1}{\sqrt{2 \pi}} \exp\left(-\nSn^2/2\right).
  \label{eq:eX-ginf-l0}
\end{equation}
There is a clear discrepancy between \Eqsref{eq:eX-ginf} and \eqref{eq:eX-ginf-l0}, suggesting a qualitative difference in the level crossing rate for a continuous and discontinuous pulse shape. This result is in agreement with the careful analysis in \Ref{bierme-2012}. The rate of level crossings is much higher for a process with jumps in the pulse shape (and continues to increase with the square root of $\gamma$ as $\gamma$ increases). No matter how strong the pulse overlap is, the discontinuous pulses are much more likely to trigger threshold crossings than the continuous pulses.

We further note that the average time above threshold for $\lambda \in {0,1}$ can be written as
\begin{equation}
    \lim_{\gamma \to \infty} \frac{\avT}{\td} \sqrt{\gamma}=  \sqrt{\frac{\pi}{2}}\erfc\left(\frac{\nSn}{\sqrt{2}}\right) \exp\left(\frac{\nSn^2}{2}\right).
  \label{eq:avT-ginf-l0}
\end{equation}
Just as the rate of level crossings increases without bound for increasing pulse overlap in the cases $\lambda = 0$ and $\lambda = 1$, the average time above threshold decreases with increasing $\gamma$. Thus, in the normal limit, the process is characterized by frequent threshold crossings but short excess times. In the case of a discontinuous pulse shape, the derivative of the process does not exist, and the method we have used to find the rate of threshold crossings is not valid (but still gives results in agreement with other methods). In this case, Rice's formula, \Eqref{eq:rice-avT} does not exist for the process (as $\dSn_\rms$ does not exist). Thus, the rate of pulse arrivals will always play a role in the expressions for the rate of threshold crossings and average excess times.

\subsection{The strong intermittency limit}\label{sec:excess_stat_strong_int_lim}
We will now investigate the limit of $\gamma \to 0$, where we can neglect overlap of individual pulses, such that each pulse appears as one isolated burst in realizations of the process. In this section, we will use $\Sn/\mA$ instead of the expressions in \Eqref{eq:norm-sn-alt}, to avoid $\gamma$ where possible. In the previous section, $\nSn$ approached a standard, normally distributed variable. Here, $\nSn$ approaches a random variable with infinite skewness and flatness, and the advantage of normalizing the signal to remove the dependence on $\mA$ is diminished. In the limit $\gamma \to 0$, we can find the number of threshold crossings, the average time above threshold and even the distribution of time above threshold for each up-crossing without going through the joint PDF of $\Sn$ and $\dSn$.

For non-overlapping pulses, the total number of upward crossings of the threshold must be the same as the total number of pulses with amplitude higher than the threshold value. Therefore, the total number of up-crossings can be written as
\begin{equation}
    \lim_{\gamma \to 0} \frac{\eX(\Sn)}{\gamma} = \sum_{K=0}^\infty P_K(K) \frac{K}{\gamma} \int\limits_{\Sn/\Snw_\text{max}}^\infty \text{d}A\, P_A(A) = \frac{\mean{K}}{\gamma} \int\limits_\Sn^\infty\text{d}A\, \frac{1}{\mA} \exp\left(- \frac{A}{\mA} \right) = \frac{T}{\td} \exp\left( -\frac{\Sn}{\mA} \right),
  \label{eq:eX-g0}
\end{equation}
where $\tmean{K} = T/\tw = \gamma T /\td$ and $\Snw_\text{max}$ is the largest positive value of $\Snw$. For the exponential pulse shape in \Eqref{eq:pulse}, $\Snw_\text{max} = \Snw(0) = 1$. This expression can also be reached by taking the limit $\gamma \to 0$ in either \Eqsref{eq:sn_X} or \eqref{eq:ex-l0}, suggesting that the number is the same for a continuous and a discontinuous pulse. This can be explained by the fact that each sufficiently large-amplitude  pulse triggers one crossing above the threshold, and this is independent of the pulse shape.

Using the complementary CDF from \Eqref{eq:sn_ccdf}, we have the total time above the threshold level $\Sn$ in the strong intermittency limit,
\begin{equation}
    \lim_{\gamma \to 0} \frac{1}{T}\frac{1-C_\Sn(\Sn)}{\gamma} = \lim_{\gamma \to 0} \frac{Q\left(\gamma,\Sn/\mA\right)}{\gamma} = \Gamma\left(0,\frac{\Sn}{\mA}\right).
    \label{eq:ccdf-g0}
\end{equation}
Estimating $\avT$ by $T\,(1-C_\Sn)/\eX$, given by \Eqsref{eq:eX-g0} and \Eqref{eq:ccdf-g0}, we find that the average time above threshold for each level crossing is given by
\begin{equation}
    \lim_{\gamma \to 0} \frac{1}{\td} \avT(\Sn) = \exp\left( \frac{\Sn}{\mA} \right) \Gamma\left( 0,\frac{\Sn}{\mA} \right).
  \label{eq:avT-g0}
\end{equation}
The rate of level crossings, given by \Eqref{eq:eX-g0}, and the fraction of time above threshold, given by \Eqref{eq:ccdf-g0}, both decay as $\gamma$ in the limit $\gamma \to 0$. Since the dependency of these two expressions on $\gamma$ is the same, the average time the signal spends above the threshold is independent of $\gamma$ in the strong intermittency limit.

\section{The distribution of excess times}\label{sec:PDF-edT-an}
In this section, we will investigate the PDF of the times spent above threshold. In the strong intermittency limit, there is a closed analytical expression for this distribution. In the normal limit, with $\lambda \to 0$, an analytical expression can also be found for crossings above the mean threshold value, but it depends explicitly on the intermittency parameter $\gamma$. In the following, we will use $L$ to denote the threshold value.

\subsection{The strong intermittency limit}
In this section, we will derive the PDF of the time above threshold in the case when overlap of pulses can be neglected, that is, the strong intermittency limit $\gamma \to 0$. For brevity of notation, we will not include the limit in the following. We will also assume $\Snw_\text{max} = \Snw(0) = 1$. Generalization to arbitrary $\Snw_\text{max}$ is done by replacing the threshold $L$ by $L/\Snw_\text{max}$.

\begin{figure}
    \centering
    \includegraphics[width=\textwidth]{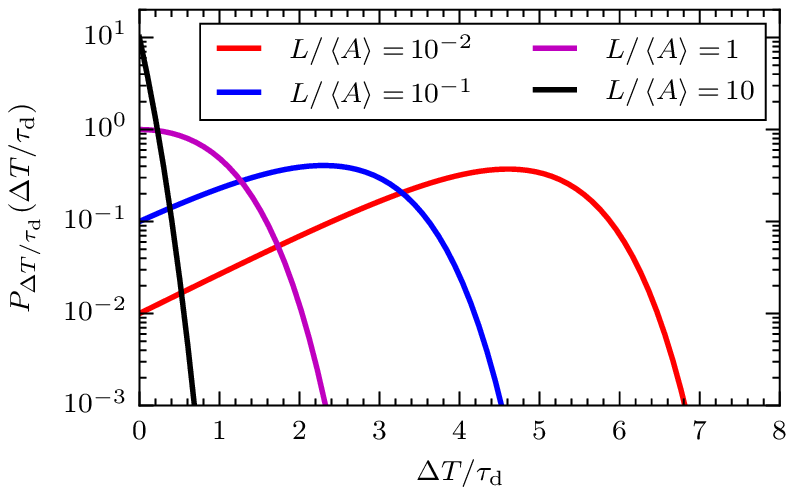}
    \caption{\label{fig:exs-pdf-DT-g0}PDF of time above threshold in the limit of strong intermittency for various threshold values.}
\end{figure}

For a given pulse with amplitude $A>L$, the signal spends a time $\edT$ above the threshold. With the two-sided exponential pulse shape, $\edT$ can be divided into a time before the peak, $\edT_{-}$, and a time following the peak, $\edT_{+}$. Assuming the pulse has peak amplitude at time $t=0$, the pulse crosses the threshold $L$ upwards at time $\edT_{-}$, given by $L = A \exp(\edT_{-}/\lambda \td)$, which gives
\begin{equation}
  \edT_{-} = - \lambda \td \ln\left( \frac{A}{L} \right).
\end{equation}
Similarly, the pulse crosses the threshold downwards at time $\edT_{+}$, given by $L = A \exp(- \edT_{+}/[(1-\lambda)\td])$, which gives
\begin{equation}
    \edT_{+} = (1-\lambda) \td \ln\left( \frac{A}{L} \right).
\end{equation}
Thus, the total time that the pulse spends above the threshold is
\begin{equation}
  \edT = \edT_{+}-\edT_{-} = \td \ln\left( \frac{A}{L} \right),
  \label{eq:excess_g0_Tn}
\end{equation}
and the pulse asymmetry plays no further role. Note that $\edT$ is always positive, since $A>L$ by assumption. Using that $A$ is exponentially distributed with mean value $\mA$, the conditional PDF of $A$ given that $A>L$, is given by the truncated exponential distribution\cite{stark-psrpe}
\begin{equation}
  P_A(A|A>L) = \frac{1}{\mA} \exp\left( -\frac{A-L}{\mA} \right),\, A>L.
  \label{eq:excess_g0_Atruc}
\end{equation}
Changing the random variable from $A$ to $\edT$ and ensuring proper normalization for the PDF of excess times gives
\begin{equation}
    P_{\edT}(\edT) =\frac{1}{\td} \frac{L}{\mA} \exp\left( \frac{1}{\td} \edT \right) \exp\left(- \frac{L}{\mA} \left[ \exp\left( \frac{1}{\td} \edT \right) -1 \right]\right),\,\edT>0.
  \label{eq:edT-g0}
\end{equation}
This is the so-called Gompertz distribution with parameters $L/(\td \mA)$ and $1/\td$, see e.g. Ref.~\onlinecite[Ch.~10.20]{olive-useful-distributions}. It is presented in \Figref{fig:exs-pdf-DT-g0} for various values of $L/\mA$. For $L\geq\mA$, the PDF decays monotonically from $\edT = 0$, while for $L<\mA$, the PDF has a maxima at $\edT/\td = \ln(\mA/L)$.
The mean value of the Gompertz distribution can be calculated as
\begin{equation}
  \avT(L) = \td \exp\left( \frac{L}{\mA} \right) \Gamma\left( 0, \frac{L}{\mA} \right),
  \label{eq:avT-g0-alt}
\end{equation}
which is equivalent to the expression in \Eqref{eq:avT-g0}. The PDF of $\edT$ will be compared to synthetic data in \Secref{sec:PDF-edT}.

\subsection{The normal limit}\label{sec:PDF-edT-norm}
It is well known that the distribution of a random variable given by a superposition of uncorrelated pulses approaches a normally distributed process in the normal limit $\gamma \to \infty$.\cite{rice-1944,garcia-prl-2012,garcia-pop-2016} In the case of a one-sided exponential pulse shape, $\lambda \to 0$, the rescaled process $\nSn$ is in the normal limit characterized by a Gaussian PDF and an exponential auto-correlation function. The statistical properties of a normally distributed random process are completely described by its PDF and auto-correlation function, and the process is thus statistically identical to any process with a standard normal distribution and exponential auto-correlation function generated by different means.

Much work has been done to elucidate the level crossing statistics and time above or below threshold for an Ornstein--Uhlenbeck (OU) process.\cite{yi-2010,alili-2005} We give the OU-process in our notation as
\begin{equation}\label{eq:OU-process}
    \rmd X(t) = - \frac{1}{\td} X(t) \rmd t  + \sqrt{\frac{2}{\td}} \rmd W(t),
\end{equation}
where $\rmd W$ is a standard Wiener process and the initial value is given by $X(0) = x_0 >0$. This process is normally distributed with mean $\tmean{X}(t) = x_0 \exp(-t/\td)$ and variance $X_\rms^2(t) = 1-\exp(2 t /\td)$, and has an exponential auto-correlation function with e-folding time $\td$. In \Appref{app:fpp-time-dependent}, it is shown that the moments of $\nSn(t)$ can be written in the same way. $X(t)$ is thus identical to $\nSn(t)$ with the conditions described above.

For the case of zero threshold, \Ref{yi-2010} gives the PDF of time above threshold $\edT$ (and a discussion of relevant references) as
\begin{equation}\label{pdf-time-above-zero-ginf}
    P_{\edT}(\edT | x_0) = \frac{x_0}{2 \td \sqrt{\pi}} \sinh(\edT/\td)^{-3/2} \exp\left(\frac{\edT}{2 \td}  - \frac{x_0^2  \exp(-\edT/\td)}{4 \sinh(\edT/\td)}\right).
\end{equation}
The initial value $x_0>0$ can be identified as the normalized value of the signal below the threshold plus the value of the pulse which brought the signal above the threshold. If the un-normalized initial value is $\Sn_0$, the relationship between $x_0$ and $\Sn_0$ is 
\begin{equation}\label{eq:sn0-to-x0}
    \Sn_0 = \rSn x_0 + \mSn.
\end{equation}
We show in \Appref{app:exp-trunc-dist} that for a threshold value $L$, $\Sn_0$ has a truncated exponential distribution
\begin{equation}
    P_{\Sn_0}(\Sn_0 | L) = \frac{1}{\mA} \exp\left(-\frac{\Sn_0 - L}{\mA}\right),\,\Sn_0>L.
\end{equation}
With $x_0$ given above and the threshold being the zero crossing of $\nSn$, which corresponds to crossing the mean value of $\Sn$ (as was also commented in \Ref{yi-2010}, crossing any stationary mean value is statistically equivalent to crossing the stationary mean value $0$), we have
\begin{equation}
    P_{x_0}(x_0) = \rSn P_{\Sn_0} (\rSn x_0 + \mSn | L = \mSn) = \sqrt{\gamma} \exp(-\sqrt{\gamma} x_0),\,x_0>0.
\end{equation}
Thus the full PDF of $\edT$ can be shown to be
\begin{align}\label{eq:pdf-edT-ginf}
    P_\edT(\edT) &= \int\limits_{0}^\infty \rmd x_0\, P_\edT(\edT | x_0) P_{x_0}(x_0) \nonumber \\
                 &= \frac{1}{\td} \sqrt{\frac{2 \gamma}{\pi}} \exp\left( \frac{2 \edT}{\td} \right) \left[ \frac{1}{\sqrt{\exp(2 \edT/\td)-1}}\right. \nonumber \\ &- \left. \sqrt{\frac{\pi \gamma}{2}} \exp\left( \frac{\gamma (\exp(2 \edT/\td)-1)}{2} \right) \erfc\left( \sqrt{\frac{\gamma (\exp(2\edT/\td)-1)}{2}} \right) \right].
\end{align}
Changing variables to $\tau = \gamma \left[ \exp(2 \edT / \td)-1 \right]/2$, this PDF can be written more compactly as 
\begin{equation}\label{eq:pdf-edT-ginf-norm}
    P_\tau(\tau) =  \frac{1}{\sqrt{\pi \tau}}- \exp(\tau) \erfc(\sqrt{\tau}),
\end{equation}
which is independent of $\gamma$. The mean value of the excess time $\edT$ can also be found,
\begin{align}\label{eq:edT-ginf-avT}
    \avT &= \int\limits_0^\infty \rmd \edT\, \edT P_{\edT}(\edT) = \frac{\td}{2} \int\limits_0^\infty \tau \ln\left( \frac{2 \tau}{\gamma} +1 \right) P_\tau(\tau) \nonumber \\ &= \frac{\td}{2} \exp\left(-\frac{\gamma}{2}\right) \left[ \pi \erfi\left( \sqrt{\frac{\gamma}{2}} \right) - \ei\left( \frac{\gamma}{2} \right) \right],
\end{align}
where $\erfi(x)=-i \erf(i x)$ and $\ei(x)$ is the exponential integral.\cite{nist-dlmf} We note that
\begin{equation}
    \lim_{\gamma \to \infty} \frac{\avT}{\td} \sqrt{\gamma} =  \sqrt{\pi/2},
\end{equation}
in agreement with the result in \Eqref{eq:avT-ginf-l0} for the threshold $\nSn = 0$. We can also find that
\begin{equation}\label{eq:edT-ginf-lim-edTinf}
    \lim_{\tau \to \infty} P_\tau(\tau) \tau^{3/2} = \frac{1}{2 \sqrt{\pi}},
\end{equation}
suggesting that $P_\edT(\edT)$ has an exponential tail for large $\edT$.

\section{Monte--Carlo studies}\label{sec:monte-carlo}

In this section, we will investigate some properties of excess time statistics for which we do not have analytical results. Firstly, we will employ a Monte--Carlo approach for investigating the PDF of $\edT$ for general $\gamma$. Secondly, the question of how quickly the rate of threshold crossings converges to the analytical value will be investigated.

\subsection{PDF of excess times}\label{sec:PDF-edT}

The PDF of excess times in the case where pulse overlap can be neglected was investigated in \Secref{sec:PDF-edT-an}, and the special case of crossings over the mean value in the case $\gamma \gg 1$ was discussed in \Secref{sec:PDF-edT-norm}. The search for an expression for the distribution of time until a process crosses a given threshold is not new, and is frequently referred to as the distribution of first passage time. The Laplace transform for the time until a FPP crosses a given threshold from below is given in \Refs{tsurui-1976,perry-2001,novikov-2005}.  The related problem of the first passage time for an Ornstein--Uhlenbeck process has been investigated by for example \Refs{madec-2004,alili-2005,yi-2010}.

To the best of the authors knowledge, there is no closed form expression for the distribution of times above threshold, and discussion of numerically computed PDFs are rare. In this section we therefore present a simulation study of the complementary CDF of $\edT$ in the case of a one-sided exponential pulse shape, $\lambda = 0$. Determining the PDF of times above threshold by simulating the process, with some examples presented in \Figref{fig:realizations}, and estimating $P_\edT(\edT)$ from the realization is computationally prohibitive, in particular for large $\gamma$ and threshold values. We will therefore use a more direct algorithm, according to the following procedure:
\begin{enumerate}
    \item At time $t=0$, a pulse arrives, taking the signal from below to above the threshold $L$. The signal takes on the value $\Sn(0) > L$ immediately after the pulse arrival. How $\Sn(0)$ is computed is discussed below. \label{itm:pulse-t0}
    \item This arrival ensures that the signal at least spends a time $t_0 = \td \ln(\Sn(0)/L)$ above the threshold, which is the excess time in the case of no other pulse arrivals in this time interval.
    \item Draw a waiting time $\tau_1$ from the exponential waiting time distribution. If $\tau_1>t_0$, the signal decays below the threshold before the next pulse arrives, and the excess time is $t_0$. If $\tau_1 < t_0$, the signal now spends a time 
        \begin{equation}
            t_1 = \td \left[ \ln \left( \Sn(0) + A_1 \exp\left(\frac{\tau_1}{\td}\right) \right) - \ln(L) \right]
        \end{equation}
        above the threshold, where $A_1$ is the exponentially distributed amplitude associated with the pulse arriving at $\tau_1$.\label{itm:compare-t1}
    \item Draw a new waiting time $\tau_2$, and compare $\tau_1 + \tau_2$ to $t_1$. If $\tau_1 + \tau_2 < t_1$, make $t_2$ in the same way as above. 
    \item Continue until the sum of the waiting times would place the arrival of the $n$'th pulse after the signal has decayed below the threshold. The time above threshold is then $t_{n-1}$ for this iteration.
    \item Repeat as often as necessary, and estimate $P_\edT(\edT)$ from all times above threshold found in steps 1-5 above.
\end{enumerate}

Step \ref{itm:pulse-t0} requires calculating $\Sn(0)$, which consists of two parts. Assume a stationary FPP takes the value $\Sn_{-}<L$ just before time $t = 0$.  A pulse with amplitude $A_0$ arrives and takes the signal above the threshold, $\Sn(0) = \Sn_{-} + A_0 > L$. It is shown in \Appref{app:exp-trunc-dist} that the PDF of $\Sn(0)$ is
\begin{equation}\label{prob:arr_above}
    P_{\Sn(0)}\left(\Sn(0)\right) = \frac{1}{\mA} \exp\left(-\frac{\Sn(0) - L}{\mA}\right),\, \Sn(0)>L,
\end{equation}
independent of the intermittency parameter $\gamma$. Samples from this distribution are readily drawn using inverse random sampling. The algorithm presented above is reasonably fast, and allows for accurate computation of the empirical CDF.

\begin{figure}[h!]
  \centering
  \begin{subfigure}{0.47\textwidth}
    \includegraphics[width = \textwidth]{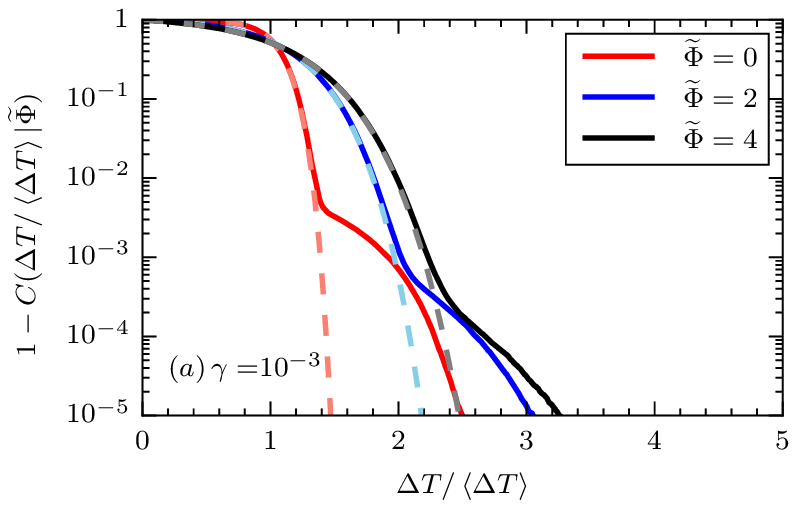}
    \caption{\label{fig:DTccdf-g0.001}}
  \end{subfigure}
  ~
  \begin{subfigure}{0.47\textwidth}
    \includegraphics[width = \textwidth]{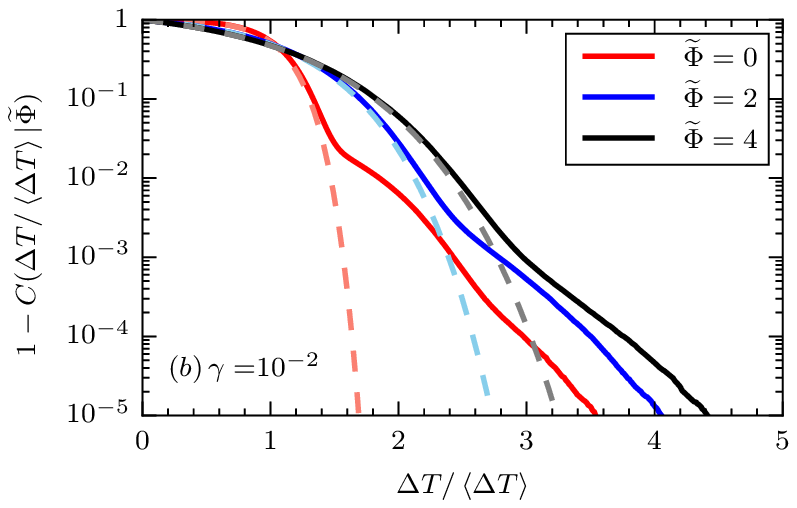}
    \caption{\label{fig:DTccdf-g0.01}}
  \end{subfigure}
  
  \begin{subfigure}{0.47\textwidth}
    \includegraphics[width = \textwidth]{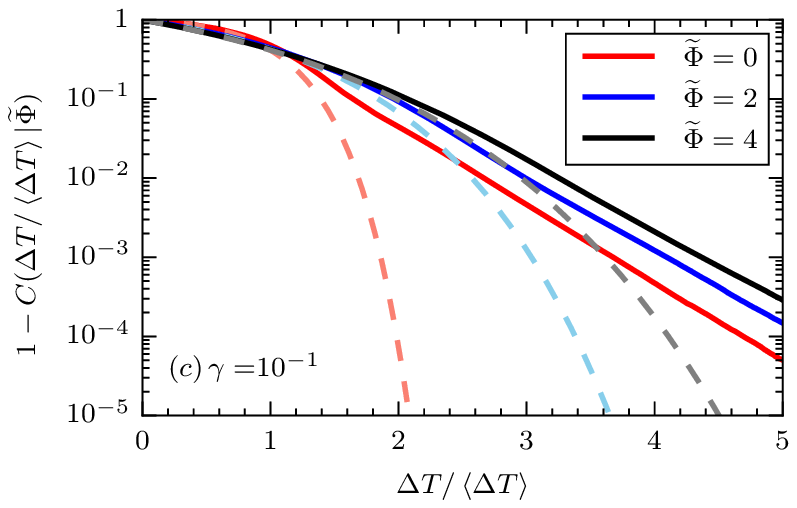}
    \caption{\label{fig:DTccdf-g0.1}}
  \end{subfigure}
  ~
  \begin{subfigure}{0.47\textwidth}
    \includegraphics[width = \textwidth]{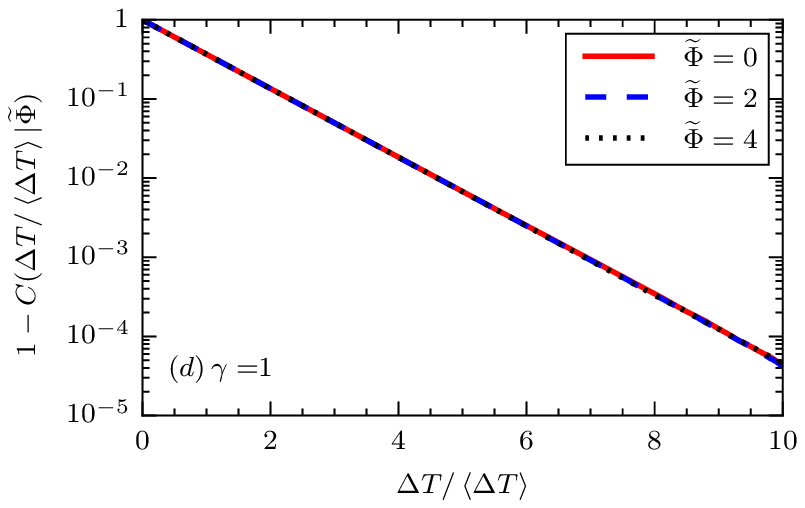}
    \caption{\label{fig:DTccdf-g1}}
  \end{subfigure}
  
  \begin{subfigure}{0.47\textwidth}
    \includegraphics[width = \textwidth]{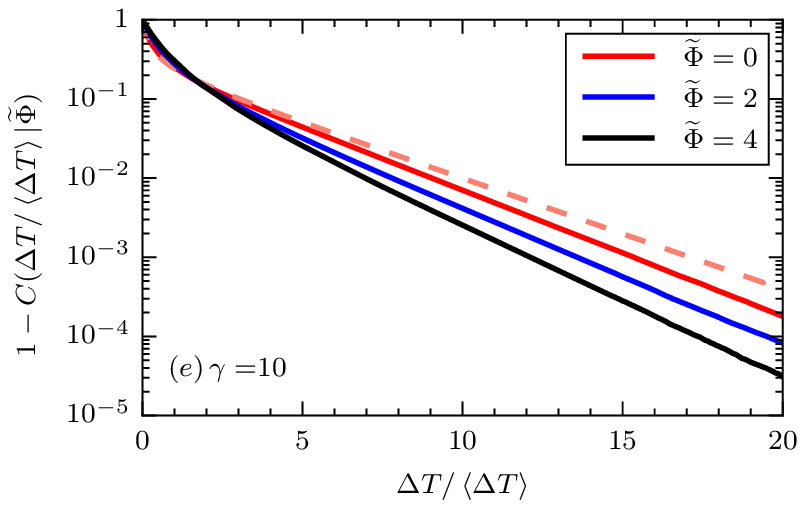}
    \caption{\label{fig:DTccdf-g10}}
  \end{subfigure}
  ~
  \begin{subfigure}{0.47\textwidth}
    \includegraphics[width = \textwidth]{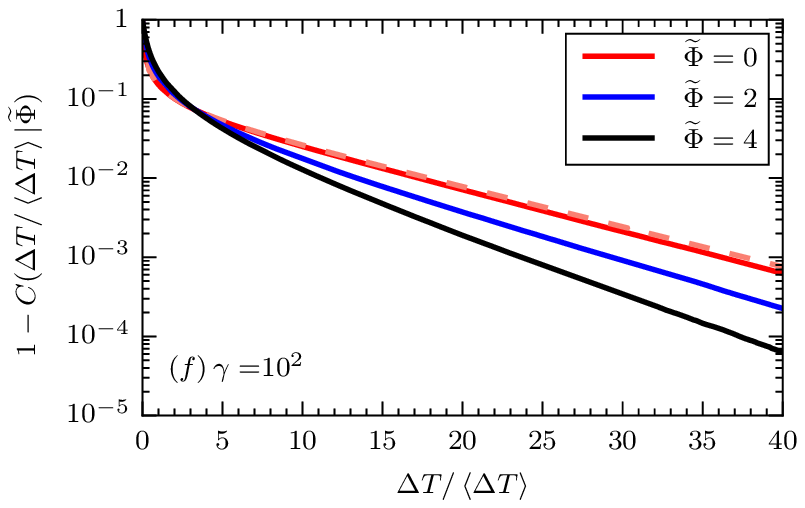}
    \caption{\label{fig:DTccdf-g100}}
  \end{subfigure}

  \caption{\label{fig:DTccdf} Synthetically generated complementary CDF of times above threshold for pulse asymmetry parameter $\lambda = 0$ and various values of the intermittency parameter $\gamma$ and threshold values. In \Figsref{fig:DTccdf-g0.001},\ref{fig:DTccdf-g0.01} and \ref{fig:DTccdf-g0.1}, the dashed lines show the analytical prediction in the limit $\gamma \to 0$. In \Figsref{fig:DTccdf-g10} and \ref{fig:DTccdf-g100}, the dashed line shows the analytical prediction for $\nSn = 0$ and $\gamma \gg 1$.}
\end{figure}

In \Figref{fig:DTccdf}, we present plots of $1-C_{\edT/\tmean{\edT}}(\edT/\tmean{\edT})$ as a function of $\edT/\tmean{\edT}$ for $\gamma \in \{10^{-3}, 10^{-2}, 10^{-1}, 1, 10, 10^{2}\}$ and various values of the rescaled threshold value $\nSn$. The full lines give the empirical complementary CDF for $10^7$ excess time simulations. In \Figsref{fig:DTccdf-g0.001}, \ref{fig:DTccdf-g0.01} and \ref{fig:DTccdf-g0.1}, the dashed lines give the complementary CDF for $\edT$ in the limit $\gamma \to 0$ given by \Eqref{eq:edT-g0}. This expression matches the simulated results for short times above threshold, but underestimate the result for longer excess times. This is due to the fact that for small but finite $\gamma$, pulse overlap is significant enough to make longer times above threshold more likely. There is a clear bump in the complementary CDF for $\gamma = 10^{-3}$, which is also visible for $\gamma = 10^{-2}$. This bump signifies the departure of the simulated distribution from the analytic result in the limit $\gamma \to 0$, and is due to the breakdown of the assumption of negligible pulse overlap, caused by the arrival of a second pulse after the original one.

In \Figsref{fig:DTccdf-g10} and \ref{fig:DTccdf-g100}, the dashed line represents the complementary CDF in the case of $\gamma \gg 1$, from \Eqref{eq:pdf-edT-ginf}. This is calculated from $1-C_\tau(\tau(\edT/\avT))$ given by \Eqref{eq:pdf-edT-ginf-norm}, where $\tau(\edT/\avT) = \gamma [ \exp(2 \avT \edT /\td) -1 ]/2$ and $\avT$ is taken from \Eqref{eq:edT-ginf-avT}. The $\gamma$-values of the respective figures have been used in this calculation. It is evident that the simulated PDF approaches the analytical one in the limit $\gamma \to \infty$.

For $\gamma <1$, the distribution is concave (on the logarithmic scale) and transitions to a convex distribution for $\gamma>1$. As seen in \Figref{fig:DTccdf-g1}, the distribution for $\gamma=1$ is an exponential distribution for all values of the threshold level. Exponential tails for large $\edT$ are seen for $\gamma = 10^{-1}$ and larger. In the limit $\gamma \to \infty$, this was already suggested by \Eqref{eq:edT-ginf-lim-edTinf}. The exponential tails are not a universal trait of this PDF; the Gompertz distribution for $\edT$ in the case $\gamma \to 0$ decays as $\exp(-\exp(\edT))$.

\subsection{Convergence of excess statistics}
In this section, we will quantify how fast the rate of level crossings converges to the analytical value. The process is as follows:

\begin{enumerate}
    \item Choose the duration $T/\td$ of a realization of the process, the imtermittency parameter $\gamma$ and the pulse asymmetry parameter $\lambda$. Generate a realization of the process.
    \item Choose $N=200$ threshold values $L_n, n = 1,2,...,199,200$ evenly spaced between $\nSn = 2$ and $\nSn = 10$, and estimate the rate of level crossings $\widehat{X}_n$ for each $L_n$.
    \item Find the mean squared logarithmic error $\mathcal{E} = \frac{1}{N} \sum_{n=1}^{N} \left[ \ln(\widehat{X}_n) - \ln(X(L_n)) \right]^2$. We use the logarithmic error instead of the linear error since the rate of threshold crossings falls exponentially with increasing threshold for large threshold values, and we wish to emphasize large threshold values.
    \item Repeat as often as necessary to estimate the mean of $\mathcal{E}$ for different $T/\td$, $\gamma$ and $\lambda$.
\end{enumerate}

\begin{figure}
  \centering
  \begin{subfigure}{0.47\textwidth}
    \includegraphics[width = \textwidth]{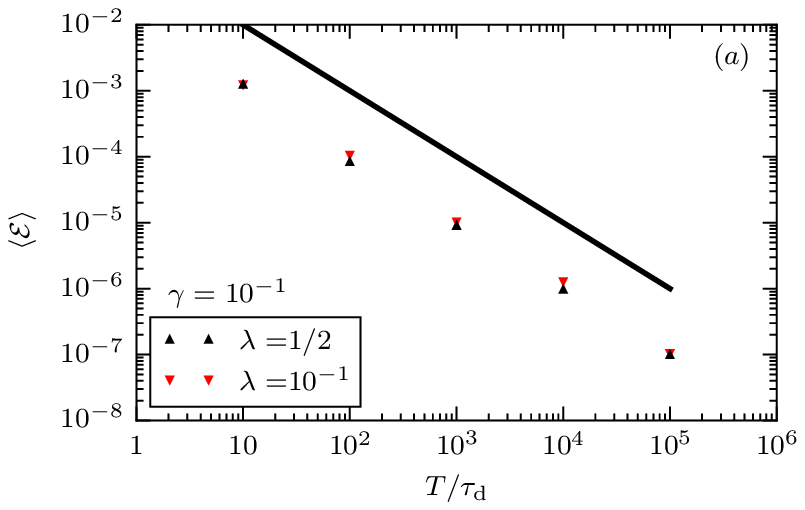}
    \caption{\label{fig:convX-g0.1}}
  \end{subfigure}
  ~
  \begin{subfigure}{0.47\textwidth}
    \includegraphics[width = \textwidth]{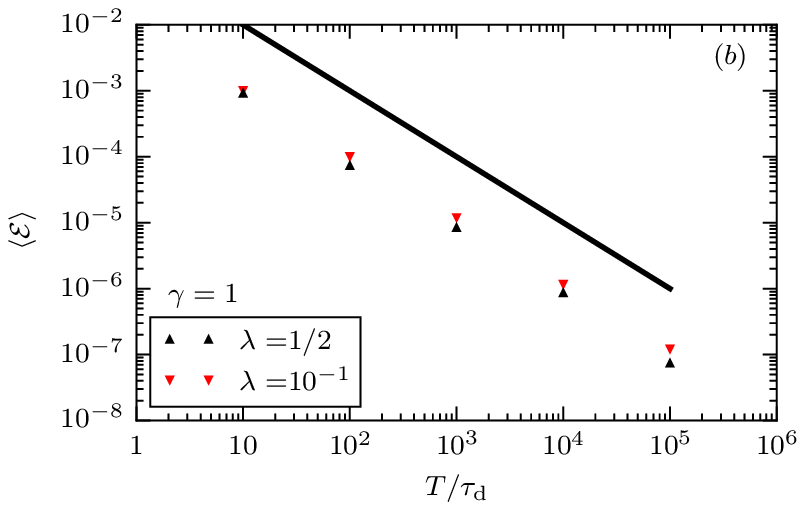}
    \caption{\label{fig:convX-g1}}
  \end{subfigure}
  
  \begin{subfigure}{0.47\textwidth}
    \includegraphics[width = \textwidth]{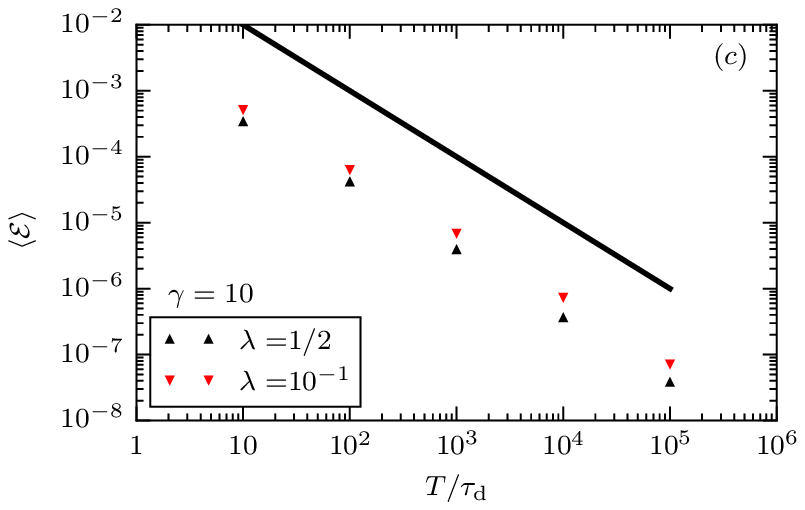}
    \caption{\label{fig:convX-g10}}
  \end{subfigure}
  ~
  \begin{subfigure}{0.47\textwidth}
    \includegraphics[width = \textwidth]{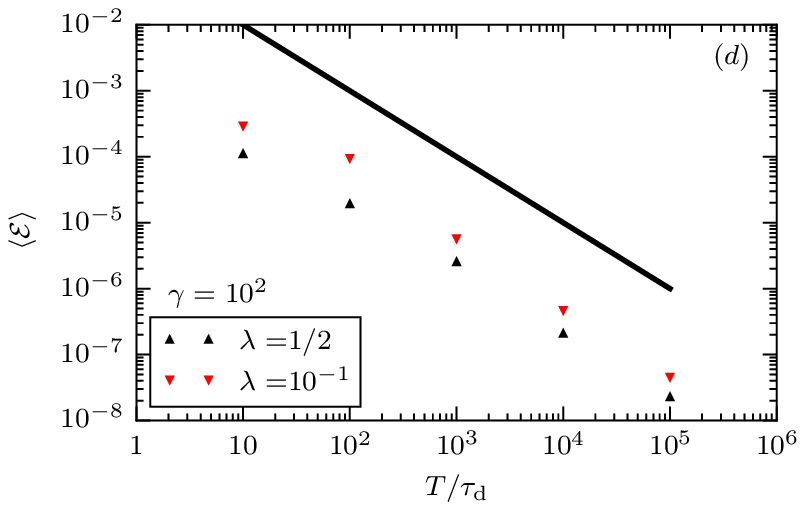}
    \caption{\label{fig:convX-g100}}
  \end{subfigure}

  \begin{subfigure}{0.47\textwidth}
    \includegraphics[width = \textwidth]{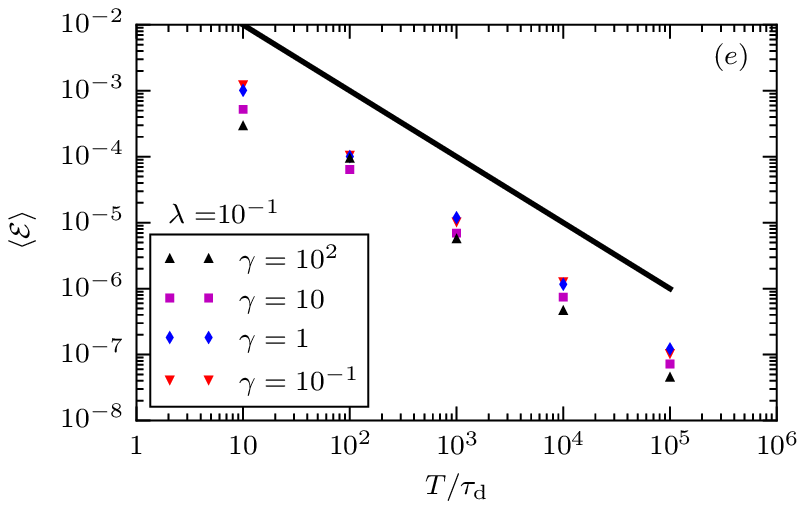}
    \caption{\label{fig:convX-l0.1}}
  \end{subfigure}
  ~
  \begin{subfigure}{0.47\textwidth}
    \includegraphics[width = \textwidth]{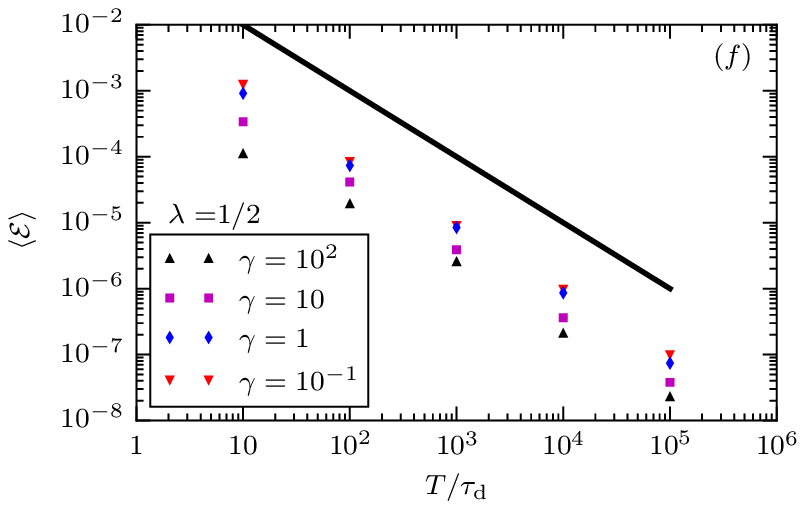}
    \caption{\label{fig:convX-l0.5}}
  \end{subfigure}

  \caption{\label{fig:convX} Mean squared error of synthetic data for various values of the pulse asymmetry parameter $\lambda$ and the intermittency parameter $\gamma$. In all cases, the solid black line gives $10^{-1} \td/T$.}
\end{figure}

In \Figref{fig:convX}, we present the estimated mean squared error of synthetic data for $\gamma \in \{ 10^{-1},1,10,10^2\}$ and $\lambda \in \{ 10^{-1},1/2\}$. The algorithm described above was repeated 100 times for each set of parameters. In all cases, the mean squared error is inversely proportional to $T/\td$. In \Figsref{fig:convX-g0.1}-\ref{fig:convX-g100}, we see that the error for  $\lambda =10^{-1}$ is larger than the error for $\lambda = 1/2$ in all cases. This is most likely a side effect of the algorithm used, where the pulses are forced to arrive at integer multiples of $\triangle_t$. This introduces a slight bias in the synthetic data, which becomes larger the more asymmetric the pulse shape is. It is also evident from \Figsref{fig:convX-l0.1} and \ref{fig:convX-l0.5} that the error decreases with increasing $\gamma$. Higher $\gamma$ for equal $T/\td$ signifies more pulses, which may lead to quicker convergence, as the samples $\{ A_k \}_{k=1}^K$ and $\{ t_k \}_{k=1}^K$ more closely reflect their underlying distributions for larger K. 

\section{Discussion and conclusion}\label{sec:conclusion}
In this contribution, a reference model for intermittent fluctuations in physical systems has been investigated. The model consists of a super-position of uncorrelated pulses with a fixed, exponential pulse shape and exponentially distributed pulse amplitudes arriving according to a Poisson process. The PDF and moments of the process were reviewed, and the moments and distribution of its derivative were discussed. The joint PDF between the process and its derivative was derived and used to obtain predictions for level crossing rates and average excess times for fluctuations above a given threshold level. These predictions depend on two model parameters, the intermittency parameter $\gamma$ and the pulse shape asymmetry parameter $\lambda$. It was shown that the functional shape of the rate of level crossings with the threshold level is strongly dependent on the intermittency parameter $\gamma$ of the process, while the functional shape of the average excess time varies little with the parameter $\gamma$. In both cases, the functional shape is independent of $\lambda$, as this parameter only appears in the pre-factor. The limit of $\lambda \to 0$ was considered, and was shown to be in agreement with previous works using different methods.\cite{bierme-2012,daly-2010} The limits of highly intermittent signals as well as the normal limit were investigated. The normal limit was shown to be in agreement with the well-known Rice's formula\cite{rice-1945} for $0<\lambda<1$, and was shown to have qualitatively different behaviour for $\lambda \in \{0,1\}$.

The PDF of the time the stochastic process spends above the threshold was found analytically in both the limit of strong intermittency for general threshold level and in the normal limit for threshold equal to the mean value, adapted from studies of Ornstein-Uhlenbeck-processes.\cite{yi-2010,alili-2005} In the strong intermittency limit, the time above threshold was shown to be Gompertz distributed. Both limits were in agreement with a Monte-Carlo study of synthetically generated time series, and the shape of the complementary CDF of time above threshold from synthetic data was presented for various values of the intermittency parameter $\gamma$.  In order to investigate the convergence of the rate of level crossings to the analytical expression, another Monte-Carlo study was performed. The convergence was shown to be proportional to $\td/T$.

The model presented here has previously been shown to be a good description of intermittent fluctuations in the boundary region of magnetically confined plasmas,\cite{garcia-pop-2013,garcia-nf-2015,kube-2016,theodorsen-ppcf-2016,garcia-nme-2016,kube-cmod,theodorsen-nf-gpi} and the rate of level crossings has compared favorably to large-amplitude fluctuations in the SOL.\cite{garcia-nme-2016,theodorsen-nf-gpi} In comparing the model to experimental data, the results presented here provide two major improvements over the classical Rice's formula in the case of intermittent fluctuations. Firstly, any discrepancy between the normal limit for excess time statistics and measurement data has previously been interpreted as a signature of intermittency in the process. The formulas derived here quantifies the level of intermittency by the model parameters $\lambda$ and $\gamma$. Secondly, Rice's formula requires the rms-value of the derivative of the signal, which is difficult if not impossible to reliably estimate for discretely sampled data containing measurement noise. In contrast, estimates for $\lambda$ and $\gamma$ can be found from the signal using the lowest order moments of $\Sn$ and its correlation function or frequency power spectrum.\cite{theodorsen-ppcf-2016,kube-2016} The variation of rate of level crossings and excess time statistics shows that the average time above threshold varies little with the intermittency parameter, suggesting that the rate of level crossings might be a more useful tool in comparing the model to experimental data in order to assess intermittency effects, although the time above threshold may be the more relevant statistic for applications in terms of failure or stability.

Even though the total time above a given threshold level may be the same for realizations of two different intermittent processes, this can be realized through either many short bursts or few but long lasting bursts events. This may have profound implications for systems where long lasting, large amplitude events can lead to severe damaging while the system can recover from the impacts of shorter burst events, depending on their frequency of occurrence. An example would be intermittent plasma-wall interactions in magnetically confined plasmas.\cite{sato-2012,fattorini-2012} Thus accurately predicting the rate of level crossings and average excess times for an intermittent process is of considerable interest to statistical modelling of fluctuations in the boundary region of magnetically confined plasmas. In future work, the novel predictions presented here will be further compared to experimental measurement data from the scrape-off layer of magnetically confined plasmas.

\section*{Acknowledgements}
This work was supported with financial subvention from the Research Council of Norway under grant 240510/F20. Discussions with M.~Rypdal are gratefully acknowledged. The authors acknowledge the generous hospitality of the {MIT} Plasma Science and Fusion Center where this work was conducted.

\appendix
\section{The joint PDF of $\Sn$ and $\dSn$ in the normal limit}\label{app:jpdf-normal-limit}
We will here demonstrate that the joint PDF of $\Sn$ and $\dSn$ given by \Eqref{jpdf} is a joint normal distribution with zero correlation coefficient in the limit $\gamma \to \infty$. We begin by changing variables to the normalized 
\begin{align}
    \nSn &= \frac{\Sn-\mSn}{\rSn}, \\
    \ndSn &= \frac{\dSn}{\dSn_\rms}.
\end{align}
where the moments of $\Sn$ and $\dSn$ are given in \Eqsref{eq:moments-sn} and \eqref{eq:moments-dsn}, respectively. Then we have 
\begin{align}
    P_{\nSn \ndSn}\left(\nSn, \ndSn \right) &= \rSn \dSn_\rms P_{\Sn \dSn}\left( \rSn \nSn + \mSn, \dSn_\rms \ndSn \right) \nonumber \\
                                            &= \frac{(\gamma \lambda)^{\gamma \lambda -1/2} \exp(-\gamma \lambda)}{\Gamma(\gamma \lambda)} \frac{[\gamma(1-\lambda)]^{\gamma(1-\lambda)-1/2}\exp(-\gamma (1-\lambda)) }{ \Gamma(\gamma(1-\lambda))} \exp(-\sqrt{\gamma}\nSn) \nonumber \\  &\times \left[ \frac{\nSn}{\sqrt{\gamma}} + \sqrt{\frac{1-\lambda}{\lambda}}\frac{\ndSn}{\sqrt{\gamma}} +1  \right]^{\gamma \lambda -1} \left[ \frac{\nSn}{\sqrt{\gamma}} - \sqrt{\frac{\lambda}{1-\lambda}}\frac{\ndSn}{\sqrt{\gamma}} +1\right]^{\gamma(1-\lambda)-1}.
\end{align}
By Stirling's formula, both fractions in the pre-factor are equal to $1/\sqrt{2 \pi}$. Using the notation
\begin{align}
    \alpha &= \nSn + \sqrt{\frac{1-\lambda}{\lambda}}\ndSn, \\
    \beta &= \nSn - \sqrt{\frac{\lambda}{1-\lambda}}\ndSn,
\end{align}
we have that $\nSn = \lambda \alpha + (1-\lambda)\beta$ and
\begin{align}\label{appeq:useful-limit}
    \lim_{\gamma \to \infty} P_{\nSn \ndSn}\left(\nSn, \ndSn \right) &= \lim_{\gamma \to \infty} \frac{1}{2\pi} \exp\left( \sqrt{\gamma} \left[ \lambda \alpha + (1-\lambda)\beta \right]\right) \left( \frac{\alpha}{\sqrt{\gamma}} +1 \right)^{\gamma \lambda -1}  \left( \frac{\beta}{\sqrt{\gamma}} +1 \right)^{\gamma(1-\lambda) -1} \nonumber \\
                                                                     &= \frac{1}{2\pi} \exp\left( \lim_{\gamma \to \infty}  \sqrt{\gamma} \left[ \lambda \alpha + (1-\lambda)\beta \right] \right. \nonumber \\ &+ \left. (\gamma \lambda -1) \ln\left( \frac{\alpha}{\sqrt{\gamma}} +1 \right)+ [\gamma(1-\lambda) -1]\ln \left( \frac{\beta}{\sqrt{\gamma}} +1 \right) \right) \nonumber \\
                                                                     &= \frac{1}{2\pi} \exp\left( \lim_{\gamma \to \infty}  \sqrt{\gamma} \left[ \lambda \alpha + (1-\lambda)\beta \right] +  (\gamma \lambda -1) \left[ \frac{\alpha}{\sqrt{\gamma}} - \frac{1}{2}\left( \frac{\alpha^2}{\sqrt{\gamma}} \right)^2 + \mathcal{O}(\gamma^{-3/2}) \right] \right. \nonumber \\ &+ \left. [\gamma(1-\lambda) -1]\left[ \frac{\beta}{\sqrt{\gamma}} - \frac{1}{2}\left( \frac{\beta^2}{\sqrt{\gamma}} \right)^2 + \mathcal{O}(\gamma^{-3/2}) \right] \right) \nonumber \\
                                                                     &= \frac{1}{2\pi} \exp\left( \lim_{\gamma \to \infty}  -\frac{\lambda}{2} \alpha^2 -\frac{1-\lambda}{2} \beta^2 + \mathcal{O}\left( \gamma^{-1/2} \right) \right) \nonumber \\
                                                                     &= \frac{1}{2\pi} \exp\left( -\frac{\nSn^2 + \ndSn^2}{2}\right).
\end{align}
Thus, in the limit of $\gamma \to \infty$, the joint PDF of $\nSn$ and $\ndSn$ approaches a joint normal distribution of two independent variables.

\section{An integral connected to the rate of threshold crossings}\label{app:half-mean-dsn}
In \Eqref{eq:X_div_PSn_independent}, the integral
\begin{equation}\label{appeq:integral-dsn}
     2 \int\limits_0^\infty \rmd \dSn\, \dSn P_\dSn(\dSn)
\end{equation}
was presented. In \Secref{sec:derivative-shot-noise}, this PDF was shown to be a convolution between a gamma distribution $P_+(\dSn)$ over positive values of $\dSn$ with shape parameter $\gamma \lambda$ and scale parameter $\mA / 2(1-\lambda)$ and a gamma distribution $P_-(\dSn)$ over negative values of $\dSn$ with shape parameter $\gamma (1-\lambda)$ and scale parameter $\mA / 2 \lambda$. The PDF of $\dSn$ is therefore
\begin{equation}
    P_\dSn(\dSn) = \int\limits_{-\infty}^{\min(\dSn,0)} \rmd x \, P_-(x) P_+(\dSn-x),
\end{equation}
where the integration limits are due to the domain of non-zero values for the gamma functions. Inserting this into \Eqref{appeq:integral-dsn} and exchanging the order of integration lets us compute the integral. The result is
\begin{multline}
    2 \int\limits_0^\infty \rmd \dSn\, \dSn P_\dSn(\dSn) = \mSn \frac{ \lambda^{\gamma \lambda} (1-\lambda)^{\gamma (1-\lambda)} \Gamma(1+\gamma)}{ \Gamma[1+\gamma(1-\lambda)]\Gamma(1+\gamma\lambda)} \times \\ \biggl\{ (1-\lambda) + \lambda\, {_2F_1}\left[1+\gamma,1;1+\gamma(1-\lambda);1-\lambda\right] - \\  \frac{\lambda (1+\gamma)(1-\lambda)}{1+\gamma(1-\lambda)} {_2F_1}\left[2+\gamma,1;2+\gamma(1-\lambda);1-\lambda\right] \biggr\} ,
\end{multline}
where $_2 F_1(a,b;c;z)$ is the hypergeometric function.\cite{nist-dlmf} This expression contains the prefactor of \Eqref{eq:X_div_PSn}, but the terms inside the curly brackets give this expression a very different behavior.

\section{The rate of level crossings in the normal limit}\label{app:normal-limit-1d}
In this Appendix, we derive a necessary result in order to go from \Eqref{eq:eX-ginf-start} to \Eqref{eq:eX-ginf}. The derivation is analogous to \Eqref{appeq:useful-limit}:
\begin{align}\label{eq:normal-limit-1d}
    \lim_{\gamma \to \infty} \left( \frac{\nSn}{\gamma^{1/2}} +1 \right)^\gamma \exp\left(-\gamma^{1/2} \nSn \right) &= \exp\left( \lim_{\gamma \to \infty} -\gamma^{1/2} \nSn + \gamma \ln\left[ \frac{\nSn}{\gamma^{1/2}} +1\right]\right) \nonumber\\
                                                                                                                     &= \exp\left( \lim_{\gamma \to \infty} -\gamma^{1/2} \nSn + \gamma \left[ \frac{\nSn}{\gamma^{1/2}} - \frac{1}{2}\left( \frac{\nSn}{\gamma^{1/2}}\right)^2 + \mathcal{O}\left(\gamma^{-3/2}\right) \right]  \right) \nonumber \\
                                                                                                                     &= \exp\left( - \frac{\nSn^2}{2}\right).
\end{align}

\section{Upper incomplete Gamma function to error function}\label{app:Q-to-erfc}
In this Appendix, an asymptotic limit of the upper incomplete Gamma function is derived. We have
\begin{equation}
    \lim_{\gamma \to \infty} Q(\gamma, \sqrt{\gamma} \nSn + \gamma) = \lim_{\gamma \to \infty} \frac{1}{\Gamma(\gamma)} \int\limits_{\sqrt{\gamma}\nSn + \gamma}^{\infty} \rmd t\, t^{\gamma-1} \exp(-t).
\end{equation}
By substituting $u = (t-\gamma)/\sqrt{\gamma}$ and using that $\gamma \Gamma(\gamma) = \Gamma(\gamma+1)$, this expression becomes
\begin{multline}
    \lim_{\gamma \to \infty} \frac{\gamma^{3/2}}{\Gamma(\gamma+1)} \int\limits_{\nSn}^{\infty} \rmd u\, (\sqrt{\gamma} u + \gamma)^{\gamma-1} \exp(-\sqrt{\gamma}u-\gamma)\\=\int\limits_{\nSn}^{\infty} \rmd u \, \lim_{\gamma \to \infty} \frac{\gamma^{\gamma+1/2}\exp(-\gamma)}{\Gamma(\gamma+1)} \left(\frac{u}{\sqrt{\gamma}} +1\right)^{\gamma-1} \exp(-\sqrt{\gamma}u).
\end{multline}
The fraction is $1/\sqrt{2 \pi}$ by Stirling's formula, and using \Eqref{eq:normal-limit-1d}, we have that
\begin{equation}
    \lim_{\gamma \to \infty} Q(\gamma, \sqrt{\gamma} \nSn + \gamma) = \frac{1}{\sqrt{2 \pi}} \int_{\nSn}^{\infty} \rmd u\, \exp\left(-\frac{u^2}{2}\right) = \frac{1}{2} \erfc\left(\frac{\nSn}{\sqrt{2}}\right).
\end{equation}
This result is used to show the equivalence between the average excess time in \Eqref{eq:sn_avT} and Rice's result in \Eqref{eq:rice-avT} in the limit $\gamma \to \infty$.

\section{Time-dependent moments of the FPP}\label{app:fpp-time-dependent}
In this Appendix, we derive the first two time-dependent moments of a normalized FPP. In \Ref{tsurui-1976}, the time-dependent characteristic function of the FPP is given as
\begin{equation}\label{eq:ch-fpp-time-dep}
    \mean{\exp(i u \Sn)} = \left( \frac{1 - i \mA \exp(-t/\td) u}{1-i \mA u} \right)^\gamma.
\end{equation}
We explicitly demand a pulse arriving at $t=0$ with value $\Sn_0$. Thus we can write a modified version of the FPP as
\begin{equation}\label{eq:FPP-initial-value}
    \Psi(t) = \Sn(t) + \Sn_0 \exp(-t/\td).
\end{equation}
We assume $\Sn_0$ is given, such that $\Psi(t)$ has the characteristic function
\begin{equation}\label{eq:ch-fpp-time-dep-2}
    \mean{\exp(i u \Psi)} = \exp\left( i u \Sn_0 \exp(-t/\td)  \right) \left( \frac{1 - i \mA \exp(-t/\td) u}{1-i \mA u} \right)^\gamma.
\end{equation}
The two first moments of $\Psi$ are $\tmean{\Psi}(t) = \gamma \mA (1-\exp(-t/\td)) + \Sn_0 \exp(-t/\td)$ and $\Psi_\rms^2(t) = \gamma \mA^2 (1-\exp(-2 t/\td))$, and the stationary moments are $\tmean{\Psi} = \mean{\Psi}(t \to \infty) = \gamma \mA$ and $\Psi_\rms^2 = \Psi_\rms^2(t \to \infty) = \gamma \mA^2$. Normalizing $\Psi$ by 
\begin{equation}
    \wt{\Psi}(t) = \frac{\Psi(t) - \mean{\Psi}}{\Psi_\rms},
\end{equation}
it is straightforward to show that 
\begin{equation}
    \mean{\exp(i u \wt{\Psi})} = \exp\left(-i \frac{\mean{\Psi}}{\Psi_\rms} u\right) \mean{\exp\left(i \frac{u}{\Psi_\rms} \Psi \right)}.
\end{equation}
Writing $\Sn_0$ as in \Eqref{eq:sn0-to-x0}, this equation can be written as
\begin{equation}\label{eq:ch-norm-fpp-time-dep}
    \mean{\exp(i u \wt{\Psi})} =\exp\left\{ i u x_0 \exp(-t/\td) + i u \sqrt{\gamma} \left[ \exp(-t/\td)-1 \right] \right\} \left( \frac{1 - i \exp(-t/\td) u/\sqrt{\gamma}}{1-i u/\sqrt{\gamma}} \right)^\gamma,
\end{equation}
whose first two moments are $\tmean{\wt{\Psi}}(t) = x_0 \exp(-t/\td)$ and $\wt{\Psi}_\rms^2(t) = 1-\exp(-2 t/\td)$. These moments are independent of $\gamma$ and are equal to the moments of the OU process in \Eqref{eq:OU-process}.

\section{The truncated exponential distribution}\label{app:exp-trunc-dist}
In this Appendix, we derive the result presented in \Eqref{prob:arr_above}. Consider a stationary stochastic process $\Sn$ consisting of a super-position of uncorrelated random pulses. Assume the pulses have a positive jump at the arrival time, and only are non-zero after the arrival time. Just before $t = 0$, the value of $\Sn$ is below the threshold $L$;
\begin{equation}\label{appeq:cond_1}
  \Sn_{-} = \lim_{\epsilon \to 0} \Sn(-\epsilon) < L.
\end{equation}
A pulse with amplitude $A$ arrives at $t = 0$, taking the signal above the threshold:
\begin{equation}\label{appeq:cond_2}
  \Sn_{-} + A = \Sn_0 > L.
\end{equation}
It is assumed that $A$ is exponentially distributed with mean value $\mA$. $\Snm$ can in principle have an arbitrary distribution.
The distribution of $\Sn_0$ is then found from integrating the joint distribution of $A$ and $\Snm$ over the region $A+\Snm < \Sn_0$, under the conditions in \Eqsref{appeq:cond_1} and \eqref{appeq:cond_2}. Since $A$ and $\Snm$ are independent, we have 
\begin{equation}
    P_{\Sn_0}(\Sn_0) = \frac{\partial}{\partial \Sn_0} C_{\Sn_0}(\Sn_0) = \frac{\partial}{\partial \Sn_0} \iint\limits_{\Sn_{-} + A < \Sn_0} \rmd A\, \rmd \Sn_{-} \, P_{\Snm}(\Sn_{-}|\Sn_{-} < L) P_A(A|A+\Sn_{-} > L),
  \label{eq:Sn0-ongoing-start}
\end{equation}
where  $C_{\Sn_0}$ is the CDF of $\Sn_0$ and
\begin{equation}
    P_{\Snm}\left(\Snm | \Snm < L\right) = \frac{P_{\Snm}\left(\Snm\right)}{C_{\Snm}(L)},\quad \Snm<L
\end{equation}
and 
\begin{equation}
    P_A\left( A | A+\Snm >L\right) = \frac{ P_A(A)}{1-C_A(L-A)},\quad A>L-\Snm.
\end{equation}
The truncated distributions are calculated by using the method given in \Ref{stark-psrpe}.
This gives
\begin{equation}
    P_{\Sn_0}\left(\Sn_0 | L \right) = \frac{\partial}{\partial \Sn_0} \int\limits_0^L \rmd \Snm \frac{P_{\Snm}\left(\Snm\right)}{C_{\Snm}(L) \left[1-C_A\left(L-\Snm\right)\right]} \int\limits_{L-\Snm}^{\Sn_0-\Snm} \rmd A\, P_A(A).
\end{equation}
The derivative with respect to $\Sn_0$ can be brought inside the first integral, and we have that
\begin{equation}
    \frac{\partial}{\partial \Sn_0} \frac{1}{1-C_A\left(L-\Snm\right)} \int\limits_{L-\Snm}^{\Sn_0-\Snm} \rmd A\, P_A(A) = P_A\left(\Sn_0-L\right),\, \Sn_0>L,
\end{equation}
where we have used that $A$ is exponentially distributed. This does not depend on $\Snm$, so we have
\begin{equation}
    P_{\Sn_0}\left(\Sn_0 | L \right) = P_A\left(\Sn_0 -L \right) \frac{ \int_0^L \rmd \Snm P_{\Snm}\left(\Snm\right)}{C_{\Snm}(L)},\, \Sn_0>L.
\end{equation}
The fraction is unity by definition, and the distribution is
\begin{equation}
    P_{\Sn_0}\left( \Sn_0 | L \right) = \frac{1}{\mA} \exp\left( - \frac{\Sn_0 -L}{\mA} \right), \Sn_0>L.
\end{equation}
Thus, if we assume exponentially distributed pulse amplitudes and pulses with jumps as described above, the distribution of the stationary process $\Sn$ plays no role for the distribution of the jumps above the threshold. In particular, for a FPP with $\lambda \to 0$ and $\gamma \to \infty$, the process is normally distributed, but the value of the signal just after the threshold is crossed is exponentially distributed.
\bibliography{crossings-sources}
\bibliographystyle{aipnum4-1} 

\end{document}